\documentclass{SciPost}

\hypersetup{
  colorlinks,
  linkcolor={red!50!black},
  citecolor={blue!50!black},
  urlcolor={blue!80!black}
}

\usepackage[bitstream-charter]{mathdesign}
\DeclareSymbolFont{usualmathcal}{OMS}{cmsy}{m}{n}
\DeclareSymbolFontAlphabet{\mathcal}{usualmathcal}

\usepackage[T1]{fontenc}
\usepackage{textcomp}
\usepackage{mathtools}
\usepackage{booktabs}
\usepackage{braket}
\usepackage{bm}
\usepackage{float}

\DeclareUnicodeCharacter{2032}{\ensuremath{\prime}}
\DeclareMathAlphabet{\mathbit}{OT1}{cmr}{bx}{it}

\fancypagestyle{SPstyle}{
  \fancyhf{}
  \lhead{\colorbox{scipostblue}{\bfseries\color{white}~SciPost Physics~}}
  \rhead{{\bfseries\color{scipostdeepblue}~Submission~}}
  
  \fancyfoot[C]{\textbf{\thepage}}
}

\begin{document}
\pagestyle{SPstyle}

\begin{center}
{\Large\textbf{\color{scipostdeepblue}{Unconventional Quantum Criticality in Long-Range Spin-1 Chains: Insights from Entanglement Entropy and Bipartite Fluctuations}}}
\end{center}

\begin{center}\textbf{
Justin Tim-Lok Chau\textsuperscript{1,2},
Jiarui Zhao\textsuperscript{3},
Nicolas Laflorencie\textsuperscript{4} and
Zi Yang Meng\textsuperscript{1$\star$}
}\end{center}

\begin{center}
    {\bfseries 1} Department of Physics and HK Institute of Quantum Science \& Technology,
    The University of Hong Kong, Pokfulam Road, Hong Kong SAR, China\\
    {\bfseries 2} State Key Laboratory of Optical Quantum Materials,
    The University of Hong Kong, Pokfulam Road, Hong Kong SAR, China\\
    {\bfseries 3} Department of Physics, The Chinese University of Hong Kong,
    Sha Tin, Hong Kong SAR, China\\
    {\bfseries 4} Univ Toulouse, CNRS, LPT, Toulouse, France\\[\baselineskip]
    $\star$ \href{mailto:zymeng@hku.hk}{\small zymeng@hku.hk}
\end{center}

\section*{\color{scipostdeepblue}{Abstract}}
\textbf{\boldmath{%
We study the ground-state phase diagram of a spin-1 Heisenberg chain with staggered long-range (LR) interactions decaying as $\propto r^{-\alpha}$ using a quantum Monte Carlo approach based on the split-spin representation. This formulation enables efficient large-scale simulations by mapping the spin-1 model onto spin-$1/2$ degrees of freedom with local projection constraints. We resolve the continuous quantum phase transition between the gapped Haldane phase at large $\alpha$ (short-range regime) and a gapless antiferromagnetically ordered N\'eel phase at small $\alpha$ (LR regime), where the continuous SU(2) symmetry is broken. From finite-size scaling and crossing point analyses, we determine the critical point to be at $\alpha_c = 2.49(1)$ and extract the associated critical exponents, which indicate unconventional criticality. In particular, the transition is found to be nonconformal, characterized by a dynamical exponent $z \neq 1$. We further analyze the scaling of entanglement entropy and bipartite fluctuations across the transition, and determine the corresponding universal scalings in both phases and at criticality.
}}

\vspace{\baselineskip}


\vspace{10pt}
\noindent\rule{\textwidth}{1pt}
\tableofcontents
\noindent\rule{\textwidth}{1pt}
\vspace{10pt}

\section{Introduction}
\label{sec:introduction}
Since the seminal work of Haldane~\cite{haldaneContinuum1983,haldaneNonlinear1983}, it is established that the low-energy physics of the one-dimensional integer-spin Heisenberg antiferromagnet is effectively described by the (1+1)-dimensional O(3) nonlinear sigma model with a topological term. For integer $S=1,2,\ldots$ the system is gapped and short-range ordered with exponentially decaying spin–spin correlations. Nevertheless, it also supports a nonlocal hidden order characterized by LR correlations in a string order parameter of topological origin~\cite{affleckRigorous1987}.
However, this relies on the assumption that interactions in the Hamiltonian are short-ranged. Interestingly, the properties of the ground state and associated phase transitions as the interaction range varies remain relatively unexplored, with only a few notable exceptions, such as studies incorporating few-body terms like biquadratic interactions or next-nearest-neighbor couplings~\cite{manmana_phase_2011,michaud_af_2012,chepiga_dimerization_2016,chepiga_comment_2016,pronk_deconfined_2025}, where the critical behavior is described by the SU(2)$_2$ WZW theory with a central charge $c=3/2$~\cite{affleck_exact_1986,affleck_critical_1987,alcaraz_conformal_1988}. 

\begin{figure}[t]
\includegraphics[width=\columnwidth]{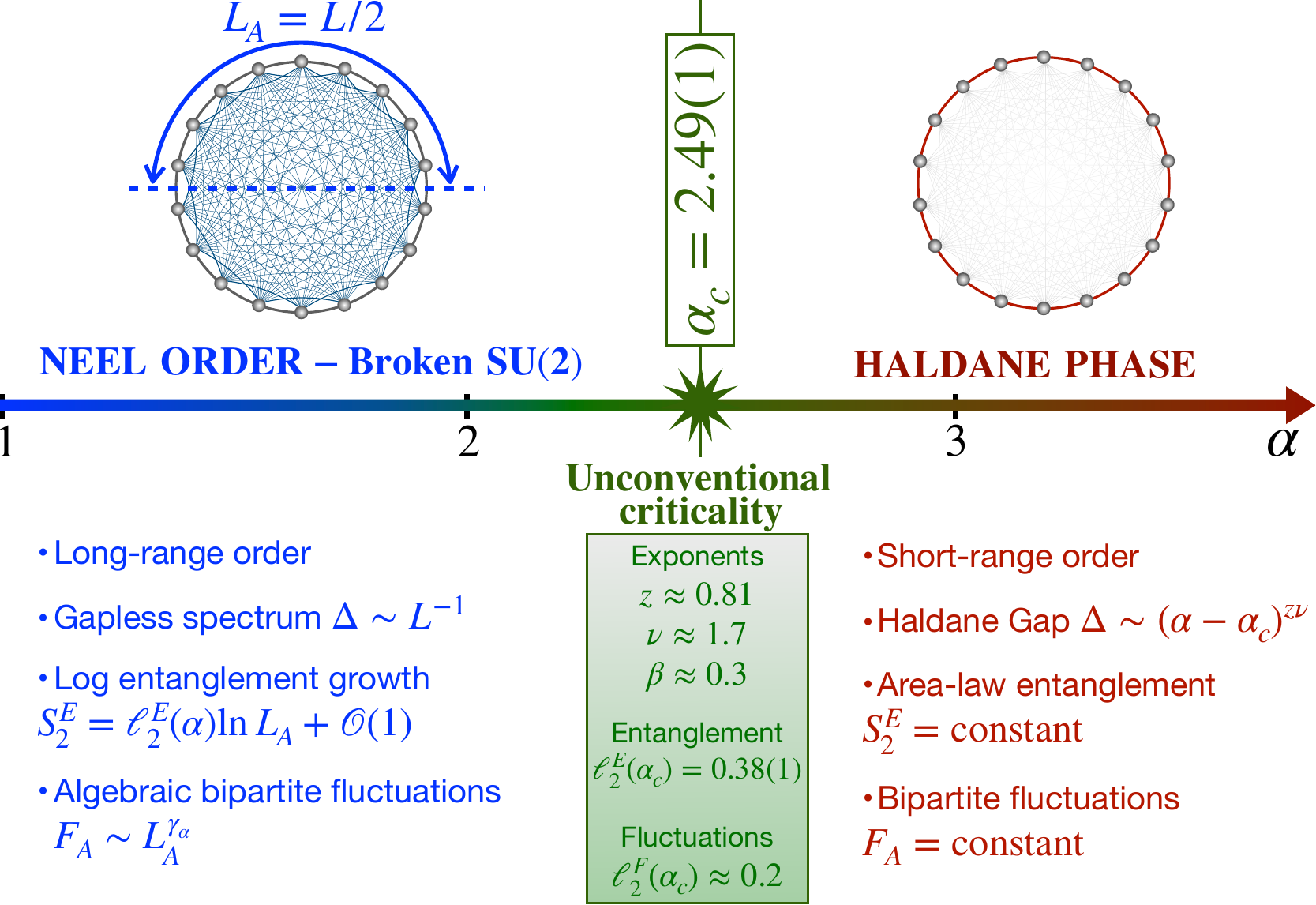}
\caption{Phase diagram and key findings. As a function of the algebraic-coupling exponent $\alpha$ in Eq.~\eqref{eq:eq1}, the LR Heisenberg $S=1$ chain exhibits two ground-state phases separated by an unconventional quantum critical point. The main results for the SU(2)-symmetry-broken N\'eel phase, the gapped Haldane phase, and the critical point at $\alpha_c = 2.49(1)$ are summarized as bullet points and discussed in the main text.}
\label{fig:fig1}
\end{figure}

In parallel, and even before these developments, pioneering studies have explored the effects of true long-range (LR) couplings, particularly the inclusion of algebraic, power-law decaying interactions in spin systems. This has remained a highly active research field for several decades, yielding key results~\cite{fisher_critical_1972,kosterlitz_phase_1976,brezin_critical_1976,cardy_one_1981,haldane_exact_88,shastry_exact_1988} (for a recent review, see~\cite{defenu_long_2023}). Among these, the celebrated Mermin-Wagner theorem~\cite{mermin_absence_1966,hohenberg_existence_1967,bruno_absence_2001,zhaoFinite2023} — later extended to zero temperature in Ref.~\cite{pitaevskii_uncertainty_1991} — stands out as a cornerstone, in particular for two- or three-dimensional systems. 
More recently, the ground state phase diagram of the LR spin chain has become an active direction of research. In particular, the fact that LR interactions can overcome the Mermin-Wagner theorem and give rise to true magnetic order breaking a continuous symmetry in 1D is now solidly confirmed based on many numerical and field theory works~\cite{laflorencieCritical2005,beach2007valencebonddescriptionlongrange,sandvik_ground_2010,maghrebi_continuous_2017,frerotEntanglement2017,vanderstraetenSpinon2020,schumm_cross_2024,zhaoUnconventional2025,yang_dynamic_2025}. However, most of these works focus on the $S=1/2$ case, aiming to characterize the phase transition between the short-range critical phase and the LR N\'eel phase, as well as the critical point separating them~\cite{zhaoUnconventional2025}.
In stark contrast, the spin-1 case remains largely uncharted, having been addressed only in a few works~\cite{gong_topological_2016,gong_kaleidoscope_2016}. Consequently, even in the seemingly simpler setting of unfrustrated LR interactions that allow continuous SU(2) symmetry breaking, both the nature of the critical point(s) and their entanglement content remain open questions.

\begin{figure}[htp!]
    \centering
    \includegraphics[width=\textwidth]{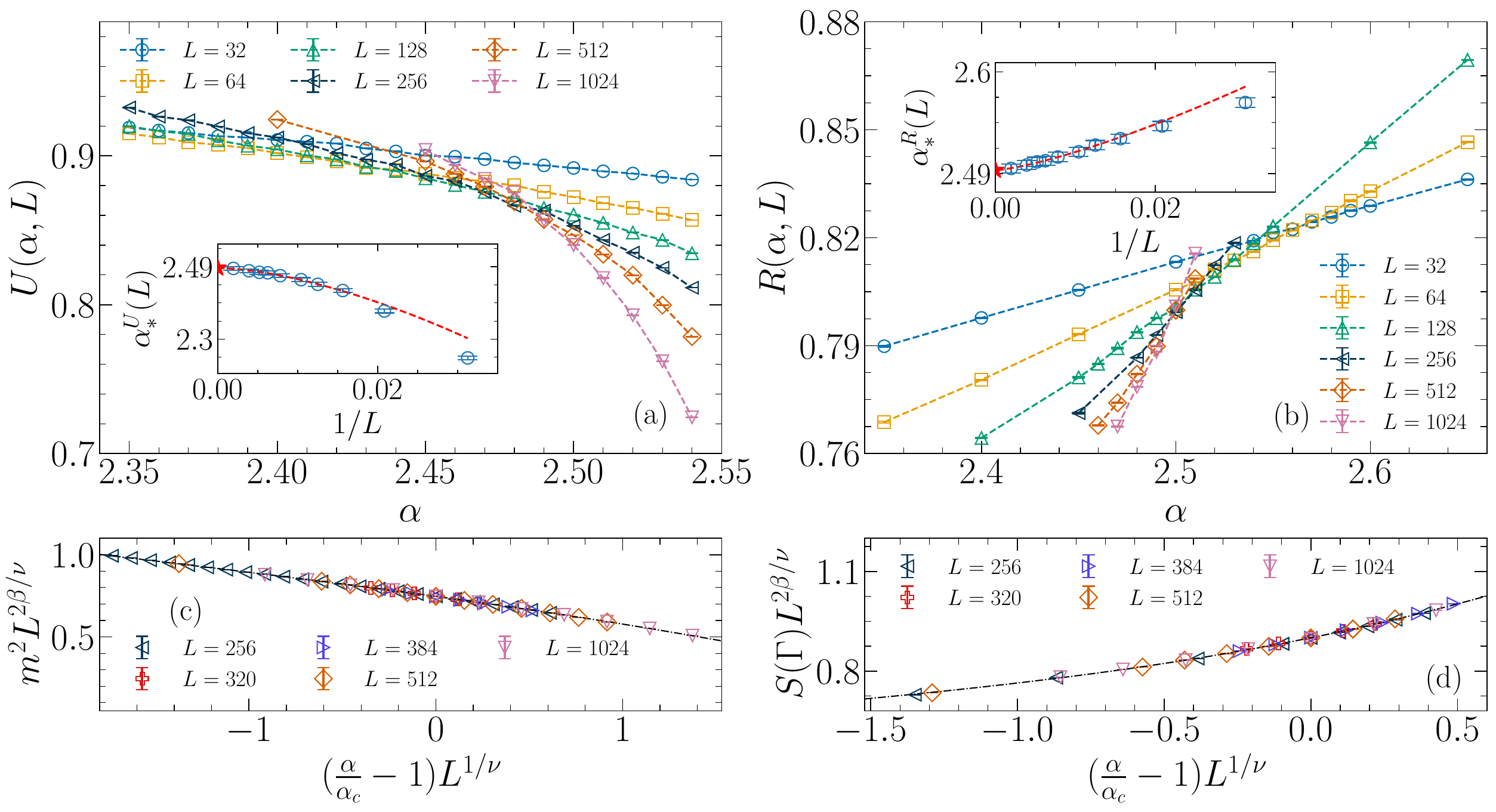}
    \caption{Determination of the QCP from crossing analysis and finite-size scaling. (a) Binder cumulant $U(\alpha,L)$ versus decay exponent $\alpha$ for system sizes $L=32$ to $1024$, with inverse temperature fixed at $\beta=L$. Inset: crossing points of $U(\alpha,L)$ and $U(\alpha,2L)$ plotted as a function of $1/L$. A power-law fit, $\alpha_*^U(L)=\alpha^{U}_{c}+bL^{-(\frac{1}{\nu} + \omega)}$, gives $\alpha^{U}_{c}=2.49(1)$. (b) SOP ratio $R(\alpha,L)$ versus $\alpha$ for the same system sizes and inverse-temperature condition. Inset: crossing points of $R(\alpha, L)$ and $R(\alpha,2L)$ versus $1/L$, fitted by the same form, yielding $\alpha^{R}_{c}=2.49(1)$. The two estimates agree within error bars, suggesting that any putative intermediate phase is restricted, if present, to at most a narrow interval of $\alpha$. We therefore identify a single critical point and use $\alpha_c = 2.49(1)$. (c) Finite-size scaling collapse of the N\'eel order parameter $m^2$, plotted as $m^2L^{2\beta/\nu}$ versus $(\alpha/\alpha_c-1)L^{1/\nu}$, using $\alpha_c = 2.49$, optimized exponents $\beta=0.27(2)$ and $\nu=1.71(9)$ are obtained and only the $L=256$, $320$, $384$, $512$ and $1024$ data are used to minimize finite-size effects. (d) Finite-size scaling collapse of $S(\Gamma)$. $S(\Gamma)L^{2\beta/\nu}$ is plotted versus the same scaling variable $(\alpha/\alpha_c-1)L^{1/\nu}$. Using the critical point $\alpha_c = 2.49$, optimized exponents $\beta=0.30(8)$ and $\nu=1.74(5)$ are obtained from the finite-size-scaling collapse. The details of extracting the optimal critical exponents $\beta$ and $\nu$ in (c) and (d) are provided in Appendix~\ref{sec:supp-fss}.}
    \label{fig:fig2}
\end{figure}

Here we determine, using quantum Monte Carlo (QMC) simulations based on the split-spin representation (Appendix~\ref{sec:supp-sse})~\cite{sandvikStochastic1999,olavQuantum2002}, the ground-state phase diagram of a spin-1 chain model with staggered (unfrustrated) LR interactions decaying as \((-1)^{r+1}/r^{\alpha}\). As shown in Fig.~\ref{fig:fig1}, we identify a quantum critical point (QCP) at \(\alpha_c=2.49(1)\)
separating the Haldane phase at large $\alpha$, with a finite spin gap and LR string order, from a gapless N\'eel-ordered phase at small $\alpha$. We observe the closing of the Haldane gap as \(\Delta\sim (\alpha-\alpha_c)^{z\nu}\) with \(z\neq 1\), showing that the transition is beyond conformal criticality. Finite-size scaling analysis of the entanglement entropy (EE) and bipartite fluctuations (BF) additionally uncover universal terms that depend continuously on \(\alpha\) within the ordered phase. At \(\alpha_c\), both EE and BF exhibit logarithmic scalings, with the entropy prefactor being interestingly close to that expected for SU(2)\(_2\) WZW.
In the Haldane phase ($\alpha>\alpha_c$), both the EE and BF follow a strict area law. In contrast, in the N\'eel-ordered phase ($\alpha<\alpha_c$), the EE increases logarithmically with $\alpha$-dependent prefactor, while the BF grows faster, as a power law. These two behaviors are consistent with linear spin-wave theory (SWT) results~\cite{frerotEntanglement2017}, at least when sufficiently far from criticality.
Possible experimental realizations are discussed at the end of this work.

\section{Results}\label{sec:results}
\subsection{Model and quantum critical point}\label{sec:model-qcp} 
The 1D LR spin-1 Heisenberg model~\cite{yusufSpin2004, laflorencieCritical2005} we study is defined by
    \begin{equation}
        H= \sum_{i<j} J_{ij} \mathbf{S}_i \cdot \mathbf{S}_j, \quad    J_{ij} = \frac{(-1)^{j-i+1}}{|j-i|^{\alpha}}.
        \label{eq:eq1}
    \end{equation}
where $\mathbf{S}_i$ is the spin-1 operator represented by a $3\times 3$ matrix, and $J_{ij}$ denotes the LR power-law interaction modified by a staggered phase factor. This model is unfrustrated and free of the sign problem, making it a suitable platform for investigating quantum fluctuations and entanglement in LR spin-1 systems using QMC simulations. Instead of directly applying the directed loop algorithm to the spin-1 Hamiltonian, we employ the split-spin representation, in which the spin-1 model is rewritten in terms of spin-1/2 Pauli matrices (see Appendix~\ref{sec:supp-sse} for details). This representation simplifies the operator structure and facilitates a more straightforward implementation of the directed loop algorithm~\cite{sandvikStochastic1999,olavQuantum2002}. 

To mitigate finite-size effects, we employ the one-dimensional Ewald summation,
\begin{equation}
\tilde{J}_{ij}=\sum_{m=-\infty}^{\infty}\frac{(-1)^{i-j+1}}{|i-j+mL|^{\alpha}},
\end{equation}
following previous studies \cite{fukuiOrder2009,koziolQuantumCritical2021,zhaoUnconventional2025,zhaoFinite2023,songQuantum2024,songDynamical2023}. With varying decay exponent $\alpha$, the system is expected to exhibit a phase transition from the N\'eel-ordered phase (small $\alpha$) to the Haldane phase (large $\alpha$). Whether an intermediate regime emerges or
whether the two phases are separated by a single critical point
is one of the questions we want to answer.

\begin{figure}[htp!]
\includegraphics[width=.95\columnwidth]{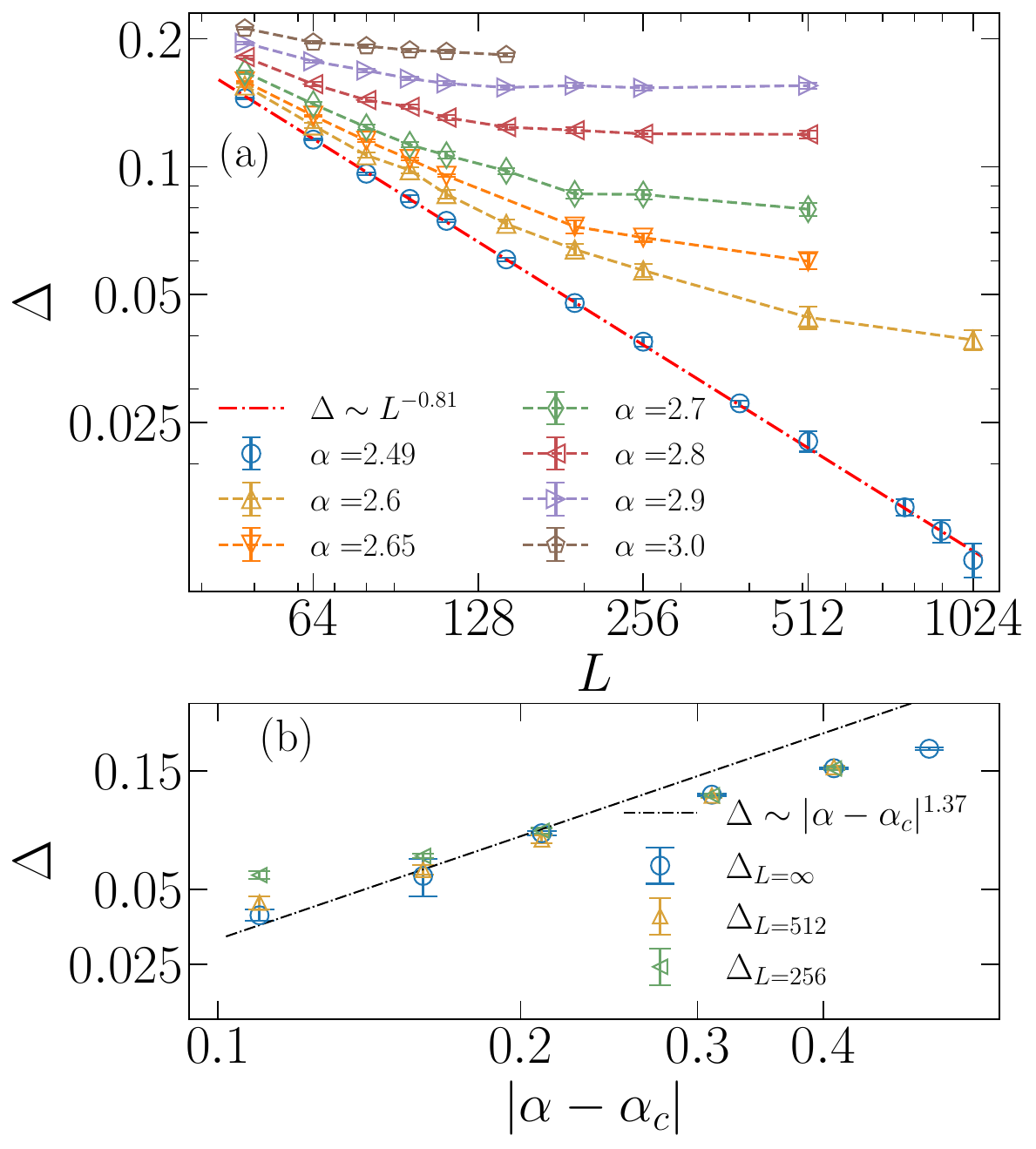}
    \caption{Finite-size scaling of the energy gap across the phase transition. (a) Energy gap $\Delta$ as a function of system size $L$ for $\alpha \in [2.49, 3.0]$. The inverse temperature is chosen as $\beta = 4L$ for $48 \le L \le 128$, $\beta = 2L$ for $L=192$, $256$, and $\beta = L$  for $L \ge 512$. Panel (a) is plotted in a log-log scale. At $\alpha_c=2.49$, a power-law fit, $\Delta \sim L^{-z}$, yields $z(\alpha_c)=0.81(1)$. (b) Thermodynamic-limit energy gap $\Delta$, obtained from fitting form $\Delta(L)=\Delta_{L=\infty} +a e^{-L/\xi}$, as a function of $|\alpha-\alpha_c|$, shown in a log-log scale. The fitting ranges are $L \in [64, 512]$ for $\alpha > 2.65$, $L \in [192, 512]$ for $\alpha \le 2.65$ and $L \in [192, 1024]$ for $\alpha=2.6$, where larger $L$ data are used to mitigate the finite-size effects. Near criticality, $\Delta$ is compared with the power law $|\alpha-\alpha_c|^{z \nu}$ with $z \nu = 1.37$, which is consistent with the data that are close to $\alpha_c$. Finite-size energy gaps $\Delta_{L=256}$ and $\Delta_{L=512}$ are included to demonstrate the finite-size effect.}
    \label{fig:fig3}
\end{figure}

To address this issue, we employ both the Binder cumulant of the N\'eel order parameter and the string-order-parameter ratio separately to locate the possible phase transition point. The dimensionless Binder cumulant for N\'eel order is defined as $U = \frac{5}{2} \left(1 - \frac{\langle m^4 \rangle}{3 \langle m^2\rangle^2}  \right)$, where $m = \frac{1}{L} \sum_{i=1}^{L} (-1)^i S_i^z$ is the N\'eel order parameter. The topological hidden order characterizing the Haldane phase in a 1D chain is captured by the string order parameter (SOP)~\cite{uedaFinite2008}, which is defined as $O^{\text{SOP}} (r) = -\langle S_0^z e^{i \pi \sum_{k=1}^{r-1} S_k^z  } S_r^z \rangle$. Correspondingly, we introduce a dimensionless quantity, the SOP ratio, as $R = 1 - \frac{S(\delta\mathbf{k})}{S(\Gamma)}$ where $\Gamma = 0$, to locate the critical point from the Haldane phase. Here $S(\mathbf{k})$ denotes the Fourier transformation of SOP measured under periodic boundary conditions, and $\delta \mathbf{k} = \frac{2 \pi}{L}$ is the smallest momentum at finite size $L$. The details of the Fourier transform of the SOP can be found in Appendix~\ref{sec:supp-sop-fourier}. According to the finite size scaling theory, the finite size crossing points $\alpha_{*}$ between $U(L)$ and $U(aL)$ or $R(L)$ and $R(aL)$ with fixed scale factor $a>1$, converge to their respective critical points as $\alpha_{*}(L) -\alpha_c \sim L^{-(\frac{1}{\nu}+\omega)}$~\cite{qinDuality2017,chenPhases2024}. Here $\nu$ is the correlation length critical exponent and $\omega$ is a non-universal correction exponent related with the leading irrelevant field in the renormalization flow~\cite{beach2005datacollapsecriticalregion}.  We therefore use this scaling relation to extract the critical points from these two quantities.

Our results of the crossing points analysis are shown in Fig.~\ref{fig:fig2}. From the data of $\alpha_{*}^{U}$ of $U(\alpha, L)$ and $U(\alpha,2L)$ in Fig.~\ref{fig:fig2} (a), we trace the crossing points in the inset and a fitting with $\alpha^{U}_{*}(L)=\alpha_c^U +b L^{-(\frac{1}{\nu}+\omega)}$ yields the extrapolated critical point $\alpha_c^U = 2.49(1)$. Independently, we also determine the $\alpha_c$ by the crossing points $\alpha_{*}^{R}$ of $R(\alpha,L)$ and $R(\alpha,2L)$ in Fig.~\ref{fig:fig2} (b). The fitting with $\alpha^{R}_{*}(L) = \alpha_c^R +b L^{-(\frac{1}{\nu}+\omega)}$ in the inset also gives $\alpha^R_c = 2.49(1)$. Our results show that the extrapolated critical points from N\'eel Binder cumulants and SOP ratios are indistinguishable within error bars, suggesting that any putative intermediate phase is restricted, if present, to at most a tiny interval of $\alpha$. This supports a single QCP  separating the N\'eel order from the Haldane phase.

Having located the critical point, we next extract the related critical exponents from finite-size scaling collapses. For the N\'eel order parameter, the scaling form $m^2L^{2\beta/\nu} \sim (\alpha/\alpha_c-1)L^{1/\nu}$ gives $\beta=0.27(2)$ and $\nu=1.71(9)$, using the few largest available system sizes to reduce the finite-size effects. A parallel collapse of the string-order structure factor with scaling form   $S(\Gamma)L^{2\beta/\nu } \sim (\alpha/\alpha_c-1)L^{1/\nu}$ yields $\beta=0.30(8)$ and $\nu=1.74(5)$, as shown in Fig.~\ref{fig:fig2} (c) and (d). Details of the procedures for extracting the critical exponents are given in Appendix~\ref{sec:supp-fss}. The two estimates are mutually consistent, and we use $\nu=1.7(1)$ in the following analysis. These obtained exponents are not of known universality in 1D QCPs, in particular, not of the Wess-Zumino-Witten SU(2)$_2$ universality class, for which $\beta=3/(4k)$ and $\nu=1/2+1/k$ with $k=2$~\cite{BELAVIN1984333}. This discrepancy motivates the dynamical scaling analysis below, where we show that the QCP is non-conformal.

\subsection{Dynamical exponent}\label{sec:dynamical-exponent}
To determine the dynamical exponent \(z\), we analyze the finite-size scaling of the low-energy gaps at the QCP. Because the Heisenberg Hamiltonian conserves the total magnetization \(S^z_{\rm tot}\), we can target fixed-\(S^z_{\rm tot}\) sectors and directly extract the corresponding excitation energies.
 At criticality, the relevant gaps obey $\Delta(L) \sim L^{-z}$. In the Haldane phase (\(\alpha>\alpha_c\)), the spectrum remains gapped and its finite-size dependence is well described by $\Delta(L)=\Delta_{L=\infty} +a e^{-L/\xi}$ where $\xi$ is the correlation length. We can thus monitor its critical closing as $\Delta \sim |\alpha -\alpha_c|^{z\nu}$, with the $z$ and $\nu$ obtained at the QCP, as displayed in Fig.~\ref{fig:fig3}. (a) shows the finite-size scaling $\Delta(L)$ for various values $\alpha\ge \alpha_c$, i.e. in the Haldane regime and at criticality, where the gap follows a power-law decay controlled by $z(\alpha_c)=0.81(1)$. This value deviates from the Lorentz-invariant case ($z=1$), showing that the LR interactions break Lorentz invariance and the QCP is not described by a conformal field theory.
 
\begin{figure}[htp!]
    \centering
    \includegraphics[width=.9\columnwidth]{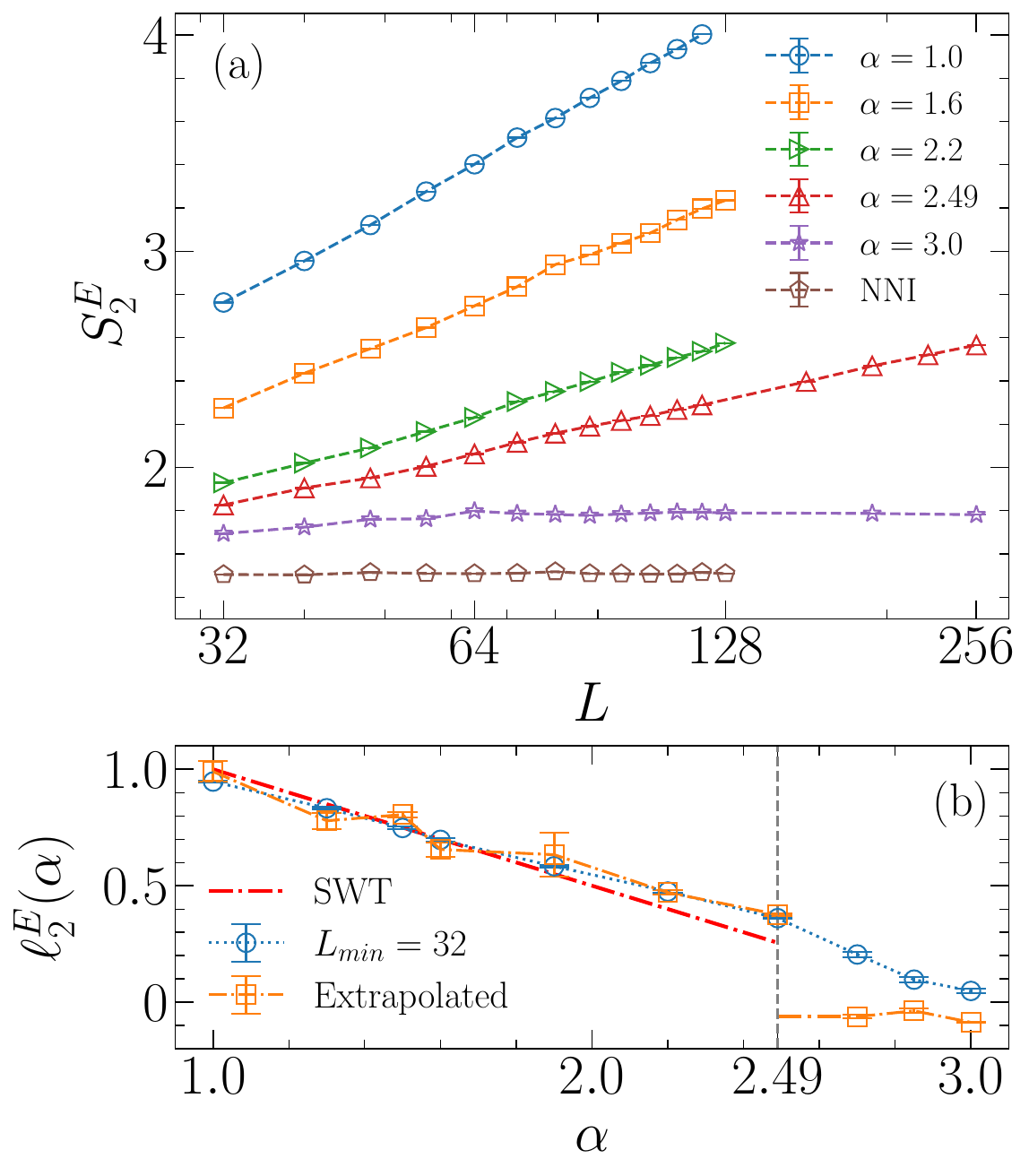}
    \caption{Half-chain R\'enyi EE scaling for different $\alpha$, with inverse temperature fixed at $\beta=L$ in the simulations. (a) $S_2^E$ as a function of system size $L$ for different values of $\alpha$. Here, NNI denotes the nearest-neighbor interaction. (b) Extrapolation of the fitted logarithmic coefficient $\ell_2^{E}(\alpha)$ to the thermodynamic limit. Blue circles correspond to results obtained with $L_{\min}=32$, while orange triangles denote the extrapolated values at the thermodynamic limit. Details of the extrapolation are provided in Appendix~\ref{sec:supp-extrapolation}. The red dash-dotted line shows the prediction from spin-wave theory (SWT), for which $S = n_G (d- z) / 2 \ln L$, where $z = (\alpha - d)/2$ and $n_G = 2$ \cite{frerotEntanglement2017}. }
    \label{fig:fig4}
\end{figure}

Fig.~\ref{fig:fig3}. (b) illustrates the closing of the gap in the Haldane phase as the system approaches the QCP. When $\alpha\to \alpha_c$, finite-size effects become increasingly pronounced due to the divergence of the correlation length, requiring larger system sizes to reliably capture the convergence of the gap. This trend is evident in Fig.~\ref{fig:fig3} (b): for $|\alpha - \alpha_c | > 0.2$, the data for $L = 256$ and $512$ are already well converged to their thermodynamic-limit values, whereas for $|\alpha - \alpha_c| \sim 0.1$, an explicit extrapolation to $L \to \infty$ becomes necessary. Nevertheless, we find that $\Delta_{L=\infty}$ in Fig.~\ref{fig:fig3} (b) indeed follow the power-law form $\Delta_{L=\infty} \sim |\alpha-\alpha_c|^{z\nu}$ with $z\nu=0.81 \times 1.7 \approx 1.37$, where the correlation length exponent $\nu=1.7(1)$ is taken from the data collapse in Fig.~\ref{fig:fig2}.

\subsection{Entanglement entropy}\label{sec:entanglement-entropy}
The $n^{\rm th}$ R\'enyi EE is defined as $S^E_n = \frac{1}{1-n} \ln \left( \text{Tr} \rho^n_A \right)$. Based on its trace structure \cite{calabreseEntanglement2004,laflorencieQuantum2016}, this is equivalent to writing it as $S^E_n  = \frac{1}{1-n}  \ln \left( \frac{Z_A^{n}}{Z^{n}}\right)$, within the QMC framework, $Z_A^{n}$ can be interpreted as $n$ replicated space-time configurations where the entangled region $A$ is connected along the imaginary time direction, while $Z^n$ corresponds to $n$ independent replicas. In this work, we focus on the second R\'enyi entropy ($n=2$) with the entanglement region $L_A=L/2$, and we employ the non-equilibrium incremental algorithm for the EE computation~\cite{demidioEntanglement2020,zhaoMeasuring2022,zhaoUnconventional2025}.

In Fig.~\ref{fig:fig4} (a), we present our QMC EE results for $\alpha \in [1,3]$, together with the nearest-neighbor interaction (NNI) case ($\alpha=\infty$). In the short-range regime, $\alpha > \alpha_c$, the system is gapped and the EE follows a strict area law $S \sim \mathcal{O}(1)$, consistent with Hastings theorem~\cite{Hastings_2007}. On the other side of the transition, in the N\'eel ordered regime  $\alpha < \alpha_c$, the measured $S^E_2$ is well-described by a logarithmic scaling
\begin{equation}
S^{E}_2 = \ell^{E}_{2} (\alpha) \ln L + \mathcal{O}(1)
\label{eq:eq2}
\end{equation}
with a prefactor $ \ell^{E}_{2}(\alpha)$ that varies continuously  with $\alpha$, similarly to the $S=1/2$ case~\cite{zhaoUnconventional2025}\footnote{We also note that in the \textit{super-long-range} regime $\alpha \le 1$, the entanglement entropy follows $S_A^2 = \ln L + \mathcal{O}(1)$ due to the tower-of-states structure~\cite{vidalEntanglement2007,liebOrdering1962}.}.
Fitting to the QMC data is performed according to Eq.~\eqref{eq:eq2} to extract the logarithmic coefficient $ \ell^{E}_{2}(\alpha)$. To mitigate finite-size effects in extracting $\ell^{E}_{2}$, we employed a fitting procedure by progressively discarding data from smaller system sizes until the coefficient value has converged
(see Appendix~\ref{sec:supp-extrapolation} for details). 

As shown in Fig.~\ref{fig:fig4} (b), for $\alpha=1$ the converged prefactor $\ell^{E}_{2}(\alpha=1) = 0.99(4)$ is very close to the expected result from the tower of states~\cite{vidalEntanglement2007}. As $\alpha$ is reduced, $\ell^{E}_{2}(\alpha)$ gradually decreases down to $ \ell^{E}_{2}(\alpha_c)=0.376(5)$ at the QCP $\alpha_c=2.49$. Very interestingly, despite the absence of Lorentz invariance ($z = 0.81 \neq 1$), we observe that the critical prefactor is remarkably close to that expected for WZW $\mathrm{SU}(2)_2$, for which one has $\ell^{E}_{2} = \frac{1}{4}\left(1+{1}/{k}\right)$. In the present case ($k=2$), this would yield $\ell^{E}_{2}(\alpha_c) = 0.375$, which is close to our numerical estimate~\footnote{We also remark that in Ref.~\cite{adelhardtUnconventional2026}, the prefactor of the critical logarithmic scaling obtained using matrix-product state calculations with open boundary conditions, for $k=1$, is also very close to the $\mathrm{SU}(2)_2$ prediction, which would be $1/4$ in their case.}. 

    For $\alpha > \alpha_c$, applying Eq.~\eqref{eq:eq2} yields $\ell^{E}_{2}(\alpha)$ values that approach zero in the thermodynamic limit. For example, at $\alpha = 2.7$, as illustrated in Fig.~\ref{fig:fig4}(b), we find $\ell^{E}_{2}(\alpha = 2.7) = 0$ upon extrapolation to the thermodynamic limit.

\begin{figure}[htp!]
\centering
\includegraphics[width=.75\columnwidth]{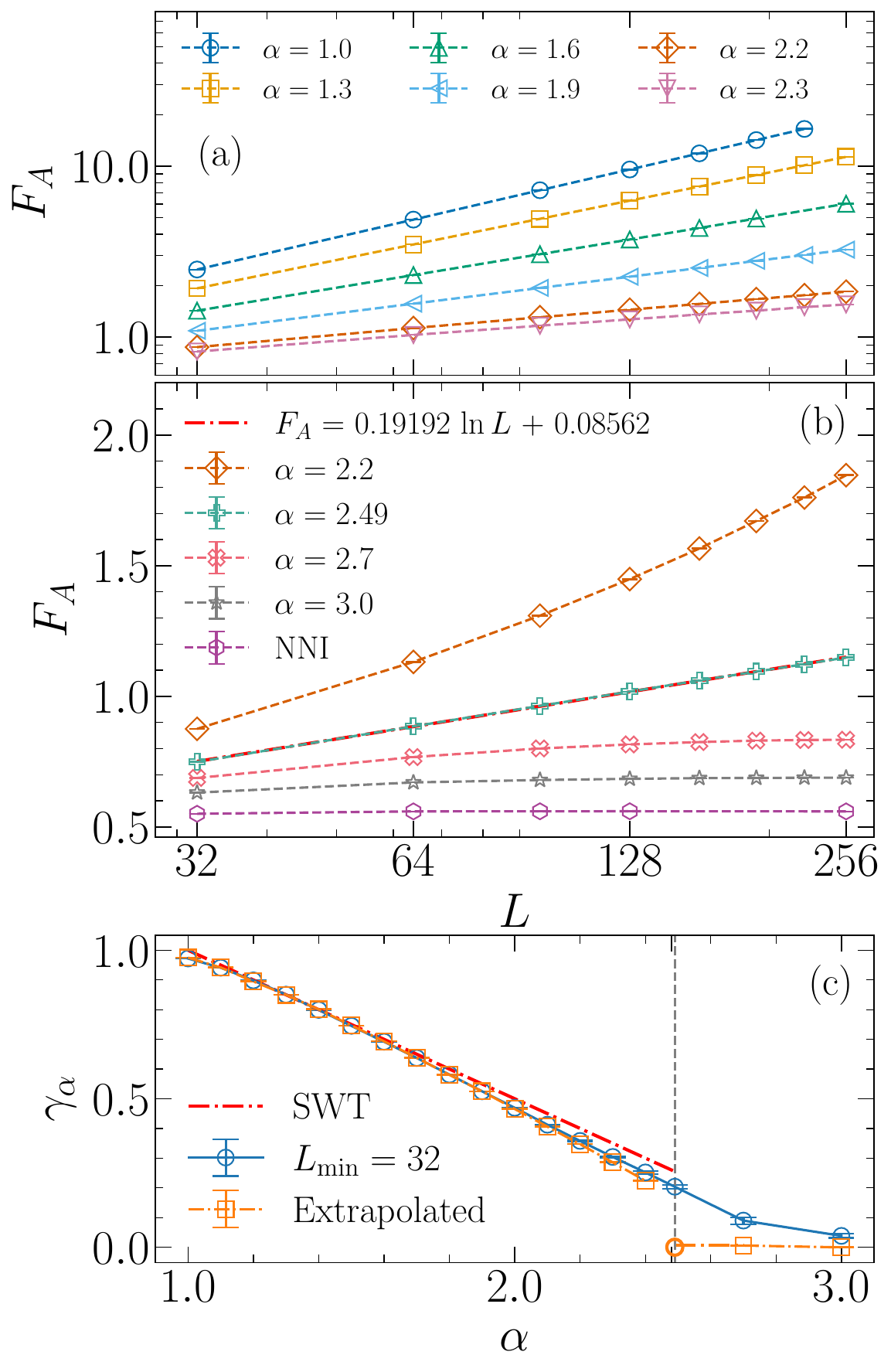}
\caption{Half-chain bipartite fluctuation for different $\alpha$, with inverse temperature fixed at $\beta =L$ in the simulations. (a) $F_{A}$ versus $L$ for $\alpha \in [1.0, 2.3]$, shown on a log-log scale. (b) $F_{A}$ versus $L$ for $\alpha \in [2.2, 3.0]$ and the nearest-neighbor interaction (NNI) case, shown on a semi-log scale. The fitted logarithmic coefficient $\ell_2^F=0.1919(8)$ at $\alpha = 2.49$ is obtained. (c) Extrapolation of the fitted power-law exponent $\gamma_{\alpha}$ to the thermodynamic limit. Blue circles are the results obtained using $L_{\min}=32$, while orange squares are the extrapolated thermodynamic limit values. {$\gamma_{\alpha_c}$, indicated by an orange circle in (c), is set to zero to indicate the vanishing of the power-law exponent at the transition. The details of the extrapolation and the justification of $\gamma_{\alpha_c} = 0$ are provided in Appendix~\ref{sec:supp-extrapolation} }. The red dash-dotted line shows the prediction from spin-wave theory (SWT), for which $F \sim L^{d-z}$ and $z = (\alpha - d)/2$ \cite{frerotEntanglement2017}.}
        \label{fig:fig5}
\end{figure}

\subsection{Bipartite fluctuations}\label{sec:bipartite-fluctuations}
 The bipartite fluctuations (BF) provide a useful diagnostic for characterizing quantum correlations and entanglement in quantum spin systems, particularly in the context of conserved quantities~\cite{klich_quantum_2009,song_bipartite_2012,rachelDetecting2012}. BF are defined as $F_A = \left\langle \left( \sum_{i \in A } O_i \right)^2 \right\rangle - \left\langle \sum_{i \in A } O_i \right\rangle^2 $, where $A$ denotes the entanglement region, and $O$ is a globally conserved quantity, which can locally fluctuate. It has been proven to be an efficient tool to detect quantum criticality with distinct scaling behaviors in different phases \cite{rachelDetecting2012}. Here, we take the total (z)-component of the magnetization, $S^z$, as the conserved quantity. 

The scaling behavior of $F_A$ for a subsystem $L_A = L/2$, reveals distinct regimes separated by $\alpha_c$. In the ordered phase $\alpha < \alpha_c$, the following power-law form  
    \begin{equation}
    F_{A} = A_\alpha L^{\gamma_{\alpha}} ,
    \label{eq:eq3}
    \end{equation}
describes the data very well, in agreement with SWT results~\cite{frerotEntanglement2017}. This clearly indicates the presence of extended correlations and nontrivial critical fluctuations. The corresponding numerical results for $F_{A}(L)$ are shown in Fig.~\ref{fig:fig5} (a), and the extracted power-law exponents $\gamma_{\alpha}$ are plotted in Fig.~\ref{fig:fig5} (c). The SWT prediction $\gamma_{\alpha}^{\rm SW}=(3-\alpha)/2$~\cite{frerotEntanglement2017}, also shown for comparison, is good deep in the ordered regime (similarly to the EE, see Fig.~\ref{fig:fig4} (b)). However, the agreement gradually deteriorates as one approaches the critical point, where at $\alpha_c$ the scaling form crosses over to a logarithmic dependence
    \begin{equation}
    F_{A} = \ell_{2}^{F} \ln L + \mathcal{O}(1),
    \label{eq:eq4}
    \end{equation}
with the fitted coefficient $\ell_{2}^{F} = 0.1919(8)$, as displayed in Fig.~\ref{fig:fig5} (b). In contrast, in the gapped Haldane phase $\alpha > \alpha_c$,  $F_{A}$ saturates to a constant at large size, reminiscent of an area law. This behavior is also evident in Fig.~\ref{fig:fig5} (c), where the extracted values no longer exhibit growth with system size.

\section{Conclusion}\label{sec:conclusion}
In this work, we present the ground-state phase diagram of a spin-1 Heisenberg chain with unfrustrated LR interactions decaying as $1/r^{\alpha}$ using a QMC approach based on the split-spin representation. A unique unconventional (non-conformal) QCP between the gapped Haldane phase at large $\alpha$ and a gapless N\'eel phase at small $\alpha$ is discovered, and the critical point
is determined to be $\alpha_c = 2.49(1)$ with the associated critical exponents. The large size scalings of both entanglement entropy and bipartite fluctuation are measured across this transition with the corresponding
universal coefficients in both phases and at the critical point revealed. We foresee the implementation of the LR interaction 1D system in programmable quantum simulator platforms such as Rydberg atom array~\cite{moegerleSpin2025} and trapped ion platforms, which allow tunable power-law interactions over a wide range of decay exponents~\cite{katzFloquet2025}. We note that continuous symmetry breaking in spin chains with LR interactions has recently been realized in the former~\cite{fengContinuous2023}.

\section*{Acknowledgements}
We thank M. Berciu, N. Chepiga, Y. D. Liao and A. Sandvik for discussions.
We thank the HPC2021 system operated by Information Technology Services at The University of Hong Kong~\cite{hpc2021}, as well as Beijing Paratera Tech Co., Ltd.~\cite{paratera}, for providing computational resources used in this work.

\paragraph{Funding information}
JTLC and ZYM acknowledge support from the Research Grants Council (RGC) of Hong Kong (Project Nos.~C7037-22GF, 17302223, 17301924 and 17301725), the ANR/RGC Joint Research Scheme sponsored by the RGC of Hong Kong and the French National Research Agency (Project No.~A\_HKU703/22), and the State Key Laboratory of Optical Quantum Materials at HKU. JRZ acknowledges support from the Research Grants Council of the Hong Kong Special Administrative Region of China through the General Research Fund (Project No.~14302725), and from the Science Panel of The Chinese University of Hong Kong through the Direct Grant scheme (Project No.~4053733). NL acknowledges partial support from the ANR research grant ManyBodyNet (No.~ANR-24-CE30-5851).

\paragraph{Use of artificial-intelligence-assisted tools}
During manuscript preparation, the authors used ChatGPT-5.5 and ChatGPT-5.6 to assist with language editing, phrasing and refinement of the presentation. The authors reviewed and edited all AI-assisted text and take full responsibility for the content of the manuscript.

\paragraph{Data and code availability}
The data supporting the findings of this study and the numerical codes used in this work are available from the corresponding author upon reasonable request.

\paragraph{Competing interests}
The authors declare no competing interests.

\paragraph{Note added}
Upon completion of this work, we became aware of a related study~\cite{adelhardtUnconventional2026} addressing an enlarged phase diagram with single-ion anisotropy using matrix-product-state calculations complemented by high-order series expansions. The results are consistent apart from the correlation-length exponent $\nu$.

\begin{appendix}
\numberwithin{equation}{section}
\numberwithin{figure}{section}
\numberwithin{table}{section}
\section{Details of the SSE QMC implementation}
\label{sec:supp-sse}

\subsection{Split-spin representation}

We use the split-spin representation~\cite{kawashima1996cluster} for spin-$S$ systems, with
$S=1/2,1,3/2,\ldots$, in which each spin-$S$ degree of freedom is
represented by $2S$ auxiliary spin-$1/2$ degrees of freedom. The
physical local Hilbert space is obtained by projecting the enlarged
$2^{2S}$-dimensional Hilbert space onto the fully symmetric
total-spin-$S$ sector. Within this projected subspace, the original
spin-$S$ operators can be written as sums of spin-$1/2$ operators,
allowing standard spin-$1/2$ Monte Carlo updates to be generalized to
higher-spin systems.

For concreteness, we illustrate the construction for the spin-1
antiferromagnetic Heisenberg model and then describe how the stochastic series expansion (SSE) quantum Monte Carlo (QMC) methods~\cite{sandvikStochastic1999,olavQuantum2002} are modified in the split-spin basis. We take the
physical Hamiltonian to be
\begin{equation}
\label{eq:spin1-H}
\mathcal H=\sum_{\langle i,j\rangle}\mathbf S_i\cdot\mathbf S_j ,
\end{equation}
where $\langle i,j\rangle$ denotes nearest-neighbor sites, $\mathbf S_i$
is a spin-1 operator, and the exchange coupling has been set to unity.

\begin{figure}[htbp]
\centering
\IfFileExists{figs1.pdf}{\includegraphics[width=0.55\linewidth]{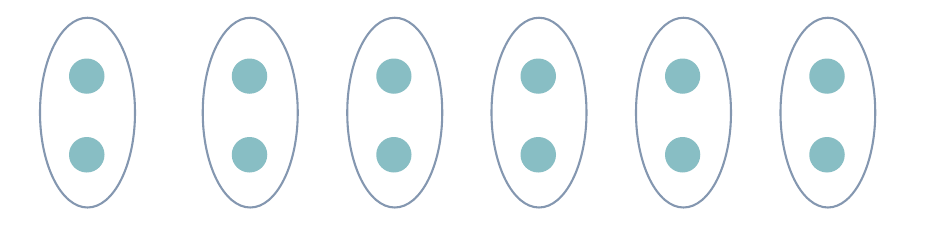}}{%
\fbox{\parbox[c][42mm][c]{0.88\linewidth}{\centering Figure file assumed available for submission: \texttt{figs1.pdf}}}}
\caption{
\textbf{Split-spin representation of a spin-1 degree of freedom.} Each filled
circle denotes an auxiliary spin-$1/2$. The two spin-$1/2$ degrees of
freedom on the same physical site are projected onto the triplet sector,
which realizes the physical spin-1 Hilbert space.
}
\label{fig:figs1}
\end{figure}

At each physical site $i$, we introduce two auxiliary spin-$1/2$
operators, $\mathbf s_{i,1}$ and $\mathbf s_{i,2}$, and identify the
spin-1 operator in the triplet sector as
$\mathbf S_i=\mathbf s_{i,1}+\mathbf s_{i,2}$, as shown in Fig.~\ref{fig:figs1}. The two-spin Hilbert
space contains one singlet state,
\begin{equation}
|0,0\rangle=\frac{1}{\sqrt2}
\left(|\uparrow\downarrow\rangle-|\downarrow\uparrow\rangle\right),
\end{equation}
and three triplet states,
\begin{equation}
\begin{aligned}
|1,1\rangle &= |\uparrow\uparrow\rangle,\\
|1,0\rangle &= \frac{1}{\sqrt2}
\left(|\uparrow\downarrow\rangle+|\downarrow\uparrow\rangle\right),\\
|1,-1\rangle &= |\downarrow\downarrow\rangle .
\end{aligned}
\end{equation}
The physical spin-1 subspace is obtained by removing the singlet
component and retaining only the triplet sector. The corresponding local
triplet projector is
\begin{equation}
P_i=\frac34+\mathbf s_{i,1}\cdot\mathbf s_{i,2}.
\end{equation}
Indeed, $P_i$ has eigenvalue one on the triplet states and zero on the
singlet state.

Substituting $\mathbf S_i=\mathbf s_{i,1}+\mathbf s_{i,2}$ into
Equation~(\ref{eq:spin1-H}) gives the split-spin form of the physical
Hamiltonian,
\begin{equation}
\mathcal H_{\rm split}
=
\sum_{\langle i,j\rangle}\sum_{\mu,\nu=1}^{2}
\mathbf s_{i,\mu}\cdot\mathbf s_{j,\nu}.
\end{equation}
In the projected subspace, $\mathcal H_{\rm split}$ is equivalent to the
original spin-1 Hamiltonian $\mathcal H$. For SSE QMC sampling it is
convenient to introduce the shifted operator
\begin{equation}
\label{eq:shifted-H}
\widetilde H
=
\sum_a H_a
=
\sum_a\left(H_{1,a}-H_{2,a}\right)
=
C N_b-\mathcal H_{\rm split},
\end{equation}
where $a$ labels an intersite bond between two auxiliary spin-$1/2$
degrees of freedom, and $N_b$ is the total number of such split-spin
bonds. The constant shift $C N_b$ does not affect normalized expectation
values but makes the diagonal matrix elements non-negative.

The physical partition function can therefore be evaluated as a trace
over the enlarged split-spin Hilbert space,
\begin{equation}
\label{eq:partition_projected}
Z_{\rm phys}
=
\operatorname{Tr}_{\rm split}
\left[
e^{-\beta \mathcal H_{\rm split}}
\prod_{i=1}^N P_i
\right]
=
e^{-\beta C N_b}
\operatorname{Tr}_{\rm split}
\left[
e^{\beta\widetilde H}
\prod_{i=1}^N P_i
\right].
\end{equation}
Equivalently, using the complete product basis
$|\alpha\rangle=|\pm\frac12,\pm\frac12,\ldots\rangle$ of the $2N$
auxiliary spins,
\begin{equation}
Z_{\rm phys}
=
e^{-\beta C N_b}
\sum_\alpha
\left\langle \alpha \left|
e^{\beta\widetilde H}
\prod_{i=1}^N P_i
\right|\alpha\right\rangle .
\end{equation}
The product $\prod_i P_i$ enforces the physical spin-1 Hilbert space
by eliminating the singlet component on every site.

\subsection{Stochastic series expansion configurations}

We decompose each local shifted bond operator $H_a$ into diagonal and
off-diagonal parts,
\begin{equation}
H_a=H_{1,a}-H_{2,a},
\end{equation}
with
\begin{equation}
H_{1,a}=C-s^z_{l(a)}s^z_{r(a)},\qquad
H_{2,a}=\frac12
\left[
s^+_{l(a)}s^-_{r(a)}
+
s^-_{l(a)}s^+_{r(a)}
\right].
\end{equation}
Here $l(a)$ and $r(a)$ denote the two auxiliary spins connected by
split-spin bond $a$, and these two spins belong to different physical
spin-1 sites. Choosing $C=1/4$, the nonzero matrix elements are
\begin{equation}
\begin{aligned}
&
\langle\uparrow\downarrow|H_{1,a}|\uparrow\downarrow\rangle
=
\langle\downarrow\uparrow|H_{1,a}|\downarrow\uparrow\rangle  =
\langle\uparrow\downarrow|H_{2,a}|\downarrow\uparrow\rangle
=
\langle\downarrow\uparrow|H_{2,a}|\uparrow\downarrow\rangle
=
\frac12 .
\end{aligned}
\end{equation}
All diagonal matrix elements of $H_{1,a}$ on parallel-spin states vanish,
and $H_{2,a}$ has no nonzero matrix elements on parallel-spin states.

The local triplet projector is decomposed in the same basis as
\begin{equation}
P_i=P_{1,i}+P_{2,i},
\end{equation}
with
\begin{equation}
P_{1,i}=\frac34+s^z_{i,1}s^z_{i,2},\qquad
P_{2,i}=\frac12
(
s^+_{i,1}s^-_{i,2}
+
s^-_{i,1}s^+_{i,2}
).
\end{equation}
The nonzero diagonal matrix elements of $P_{1,i}$ are
\begin{equation}
\label{eq:Pi_diag_parallel}
\langle\uparrow\uparrow|P_{1,i}|\uparrow\uparrow\rangle
=
\langle\downarrow\downarrow|P_{1,i}|\downarrow\downarrow\rangle
=
1,
\end{equation}
and
\begin{equation}
\label{eq:Pi_diag_antiparallel}
\langle\uparrow\downarrow|P_{1,i}|\uparrow\downarrow\rangle
=
\langle\downarrow\uparrow|P_{1,i}|\downarrow\uparrow\rangle
=
\frac12 .
\end{equation}
The nonzero off-diagonal matrix elements of $P_{2,i}$ are
\begin{equation}
\label{eq:Pi_offdiag}
\langle\uparrow\downarrow|P_{2,i}|\downarrow\uparrow\rangle
=
\langle\downarrow\uparrow|P_{2,i}|\uparrow\downarrow\rangle
=
\frac12 .
\end{equation}

Expanding Equation~(\ref{eq:partition_projected}) in powers of $\beta$ gives
\begin{equation}
\label{eq:SSE_expansion}
Z_{\rm phys}
=
e^{-\beta C N_b}
\sum_\alpha
\sum_{n=0}^{\infty}
\frac{\beta^n}{n!}
\sum_{S_n}
\left\langle\alpha\left|
\prod_{p=1}^{n} O_p
\prod_{i=1}^{N}P_i
\right|\alpha\right\rangle .
\end{equation}
Here $S_n=\{O_1,O_2,\ldots,O_n\}$ denotes an ordered sequence of local
Hamiltonian vertices, with each $O_p$ chosen from the allowed diagonal
or off-diagonal bond operators. $n$ is the number of nonidentity Hamiltonian vertices. The projector operators are not
generated by the Taylor expansion; instead, one local projector is
attached to each physical site.

In simulations, the Taylor series is truncated at a sufficiently large
cutoff $M$, and identity operators are inserted so that all Hamiltonian
operator strings have fixed length $M$. The product of projectors is then
expanded as $\prod_iP_i=\prod_i(P_{1,i}+P_{2,i})$, so that each
site contributes one diagonal or off-diagonal projector vertex. The
fixed-length SSE representation can be written as
\begin{equation}
\label{eq:fixed_length_SSE}
Z_{\rm phys}
=
e^{-\beta C N_b}
\sum_{\alpha}
\sum_{S_M}
\sum_{\Gamma_N}
\frac{\beta^n(M-n)!}{M!}
\left\langle\alpha\left|
\prod_{p=1}^{M}O_p
\prod_{i=1}^{N}P_{\gamma_i,i}
\right|\alpha\right\rangle .
\end{equation}
Here \(\gamma_i=1,2\) labels the diagonal or off-diagonal part of the local
projector \(P_{\gamma_i,i}\), and \(\Gamma_N\) denotes the ordered sequence of all \(N\)
local projector vertices. These vertices form an additional fixed layer in the
linked-vertex representation used for the SSE QMC loop updates.

\begin{figure}[t]
\centering
\IfFileExists{figs2.pdf}{\includegraphics[width=0.85\linewidth]{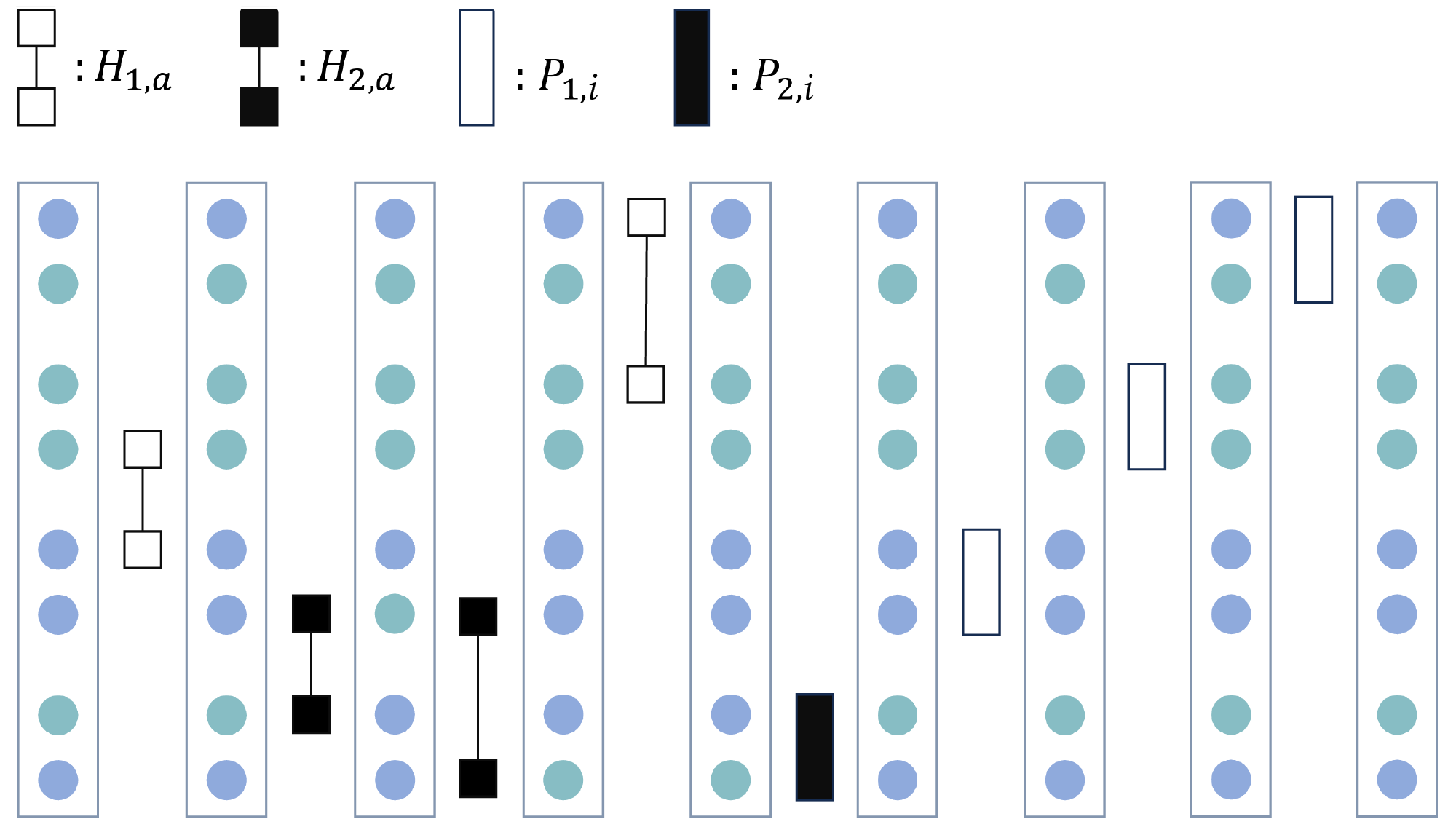}}{%
\fbox{\parbox[c][42mm][c]{0.88\linewidth}{\centering Figure file assumed available for submission: \texttt{figs2.pdf}}}}
\caption{
\textbf{SSE configuration in the split-spin representation.} The filled circles
denote the $S^z=\pm1/2$ states of the auxiliary spins. The Hamiltonian
part of the operator string contains diagonal vertices $H_{1,a}$,
off-diagonal vertices $H_{2,a}$, and identities. The projector layer
contains one local vertex, $P_{1,i}$ or $P_{2,i}$, for each physical
site. The Hamiltonian vertices act on auxiliary spins belonging to
different physical sites, whereas the projector vertices act on the two
auxiliary spins on the same physical site.
}
\label{fig:figs2}
\end{figure}

Thus the configuration space is the usual spin-$1/2$ SSE configuration
space supplemented by one projector vertex per physical spin-1 site, as illustrated in Fig.~\ref{fig:figs2}.
Apart from these additional on-site vertices, the diagonal and
directed-loop updates follow the standard spin-$1/2$ SSE construction.

\subsection{Diagonal and directed-loop updates}

The diagonal update sweeps through the $M$ positions of the Hamiltonian
operator string and attempts to replace an identity operator by a
diagonal bond operator $H_{1,a}$, or the reverse. If position $p$
contains the identity, a split-spin bond $a$ is chosen uniformly from the
$N_b$ intersite auxiliary-spin bonds. The insertion $I\rightarrow
H_{1,a}$ is accepted with probability
\begin{equation}
P_{\rm ins}
=
\min\left[
1,\,
\frac{\beta N_b
\langle\alpha(p)|H_{1,a}|\alpha(p)\rangle}
{M-n}
\right].
\end{equation}
Here $|\alpha(p)\rangle$ is the propagated basis state immediately
before position $p$, and $n$ is the current number of nonidentity
Hamiltonian vertices. Conversely, if position $p$ contains a diagonal
operator $H_{1,a}$, the removal $H_{1,a}\rightarrow I$ is accepted with
probability
\begin{equation}
P_{\rm rem}
=
\min\left[
1,\,
\frac{M-n+1}
{\beta N_b
\langle\alpha(p)|H_{1,a}|\alpha(p)\rangle}
\right].
\end{equation}
Because the number of projector vertices is fixed to $N$, no insertion
or removal updates are applied to the projector layer. A convenient
initial configuration consists of $M$ identity operators followed by $N$
diagonal projector vertices $P_{1,i}$, with the auxiliary-spin states
initialized randomly.

The directed-loop update is performed on the linked-vertex
representation of the combined Hamiltonian string and projector layer. A
loop is initiated by choosing a random leg and flipping the corresponding
auxiliary spin, thereby creating a discontinuity. The discontinuity is
propagated through the linked vertices until it returns to its starting
point and the loop closes. When the loop enters a vertex through a given
leg, the exit leg is chosen from the allowed exits using scattering
probabilities that satisfy local detailed balance,
\begin{equation}
W(v)P(v\rightarrow v')
=
W(v')P(v'\rightarrow v).
\end{equation}
Here $W(v)$ and $W(v')$ are the matrix-element weights of the initial
and final local vertex configurations.

For the Hamiltonian vertices, choosing $C=1/4$ makes all nonzero local
weights equal to $1/2$. The directed-loop equations then admit the usual
bounce-free solution of the spin-$1/2$ Heisenberg SSE algorithm. In this
solution, the loop exits through a leg that converts the incoming vertex
into another allowed vertex with the same weight, thereby interconverting
diagonal and off-diagonal Hamiltonian vertices.

For the projector layer, the possible local vertices are those listed in
Equations~(\ref{eq:Pi_diag_parallel})--(\ref{eq:Pi_offdiag}). Their weights
are $1$, $1/2$, and $1/2$, respectively, for the parallel diagonal,
antiparallel diagonal, and antiparallel off-diagonal projector vertices.
We choose the exit leg using heat-bath probabilities over the locally
allowed final vertices. If $\mathcal A(j)$ denotes the set of allowed
final vertices for an entrance into a vertex of type $j$, the transition
probability is
\begin{equation}
P(j\rightarrow i)
=
\frac{W_i}
{\sum_{k\in\mathcal A(j)}W_k},
\qquad
i\in\mathcal A(j).
\end{equation}

In summary, the split-spin spin-1 SSE algorithm differs from the
conventional spin-$1/2$ Heisenberg SSE algorithm in two ways. First, each
physical spin-1 site is represented by two auxiliary spin-$1/2$ world
lines. Second, the linked-vertex representation contains an additional
on-site projector vertex for each physical site. With these
modifications, the diagonal update and directed-loop update proceed
exactly as in the standard spin-$1/2$ SSE framework.

\section{Finite-size extrapolation of entropy and fluctuations}
\label{sec:supp-extrapolation}

\begin{figure}[htp!]
    \centering
\IfFileExists{figs3.pdf}{\includegraphics[width=0.98\linewidth]{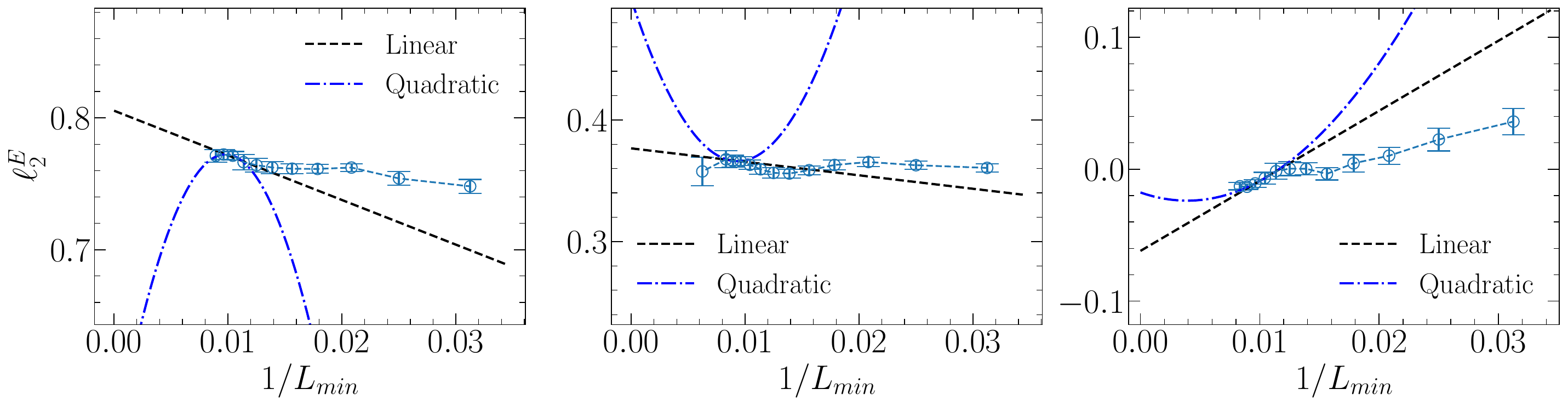}}{%
\fbox{\parbox[c][42mm][c]{0.98\linewidth}{\centering Figure file assumed available for submission: \texttt{figs3.pdf}}}}
    \caption{\textbf{Finite-size extrapolation of the entropy coefficient.} Extracted $\ell^{E}_{2}$ as a function of $1/L_{\min}$ in  N\'eel phase ($\alpha = 1.5$), at the critical point ($\alpha = 2.49$), and in the Haldane phase ($\alpha = 3.0$) (left to right). For each $L_{\min}$, the coefficient $\ell^{E}_{2}$ is obtained from fitting $S_2^E =\ell_{2}^{E}\ln L + O(1)$ over system sizes $L \ge L_{\min}$. The drift with $1/L_{\min}$ indicates finite-size corrections from the smallest sizes, while the increased scatter at large $L_{\min}$ indicates the reduced number of points in the fitting window. Dashed lines are linear fits and dash-dotted curves are quadratic fits to the selected windows $L_{\min}=88,96,104$ (N\'eel), $104,112,120$ (critical point), and $96,104,112$ (Haldane). The $1/L_{\min}\to 0$ intercept of the linear fit is used as the final estimate of $\ell^{E}_{2}$. Error bars show the standard error of the fitted slope.}
    \label{fig:figs3}
\end{figure}

Our aim is to reduce finite-size effects in the extraction of $\ell^{E}_{2}$. We therefore fit the data to
     \begin{equation}
     S_2^E =\ell_{2}^{E}(\alpha)\ln L + O(1),\qquad L_A=L/2,
     \end{equation}
and examine the fitted coefficient as a function of the lower cutoff $L_{\min}$ of the fitting window. Specifically, we successively exclude the smallest system sizes and repeat the fit until only four system sizes remain, generating a sequence of estimates for $\ell^{E}_{2}$. As shown in Fig.~\ref{fig:figs3}, the  extracted slopes in N\'eel phase ($\alpha = 1.5$), at the critical point ($\alpha = 2.49$), and Haldane phase ($\alpha = 3.0$) vary systematically with $1/L_{\min}$, indicating appreciable finite-size corrections from small-$L$ data. On the other hand, choosing too large an $L_{\min}$ leaves too few points in the fitting window and makes the fitted slopes more susceptible to fluctuations. We therefore identify a converged regime in $1/L_{\min}$ that balances these two effects.

To do so, we define the converged regime by requiring two conditions to be satisfied simultaneously: the change in slope between adjacent points is smaller than $0.02$, and the coefficient of variation of the fitted values does not exceed $0.2$. We then estimate thermodynamic-limit value of $\ell^{E}_{2}$ by a linear extrapolation in $1/L_{\min}$ using the converged point and the next two points. As shown in Fig.~\ref{fig:figs3}, this linear form provides a stable description of the converged data in all three phases, whereas higher-order fits are more sensitive to the remaining fluctuations. We therefore use the linear extrapolation to quote the final estimate of $\ell^{E}_{2}$.

We apply the same analysis to the bipartite fluctuation $\ell^{F}_{2}$. Recall that bipartite fluctuation has the following power-law scaling form in the N\'eel phase,

    \begin{equation}
        F_{A} = A_\alpha L^{\gamma_{\alpha}}
        \label{eq:FA-power-law-supp}
    \end{equation}

\begin{figure}[htp!]
    \centering
\IfFileExists{figs4.pdf}{\includegraphics[width=0.6\linewidth]{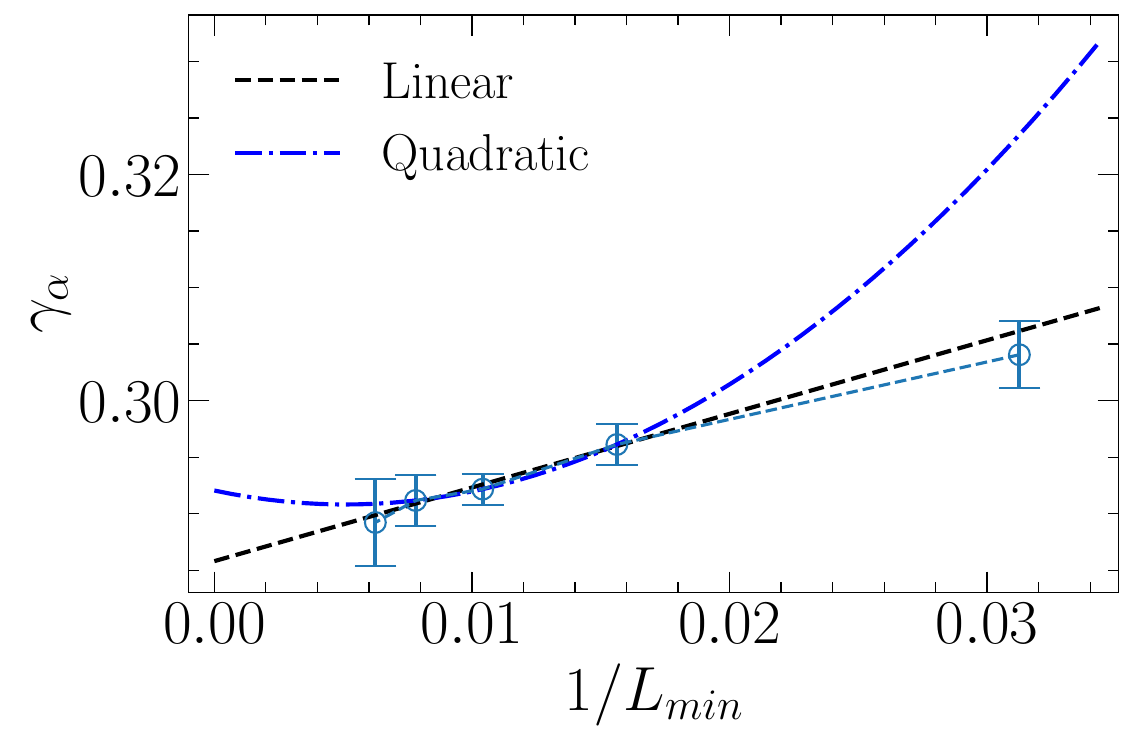}}{%
\fbox{\parbox[c][60mm][c]{columnwidth}{\centering Figure file assumed available for submission: \texttt{figs4.pdf}}}}
    \caption{\textbf{Finite-size extrapolation of the fluctuation exponent.} Extracted $\gamma_{\alpha}$ as a function of $1/L_{\min}$ in the N\'eel phase ($\alpha=2.4$). For each $L_{\min}$, the exponent $\gamma_{\alpha}$ is obtained by fitting $F_{A}=A_{\alpha}L^{\gamma_{\alpha}}$ over system sizes $L\geq L_{\min}$. The drift with $1/L_{\min}$ indicates finite-size corrections from small-$L$ data, while the increased scatter at large $L_{\min}$ indicates the reduced number of points in the fitting window. Dashed lines show linear fits and dash-dotted curves quadratic fits over the selected window $L_{\min}=64,96,128$. The $1/L_{\min}\to 0$ intercept of the linear fit is taken as the final estimate of $\gamma_{\alpha}$, following the same extrapolation procedure used for $\ell^{E}_{2}$. Error bars denote the standard error of the fitted exponent.}

    \label{fig:figs4}
\end{figure}

In direct analogy with the extraction of $\ell^{E}_{2}$, we track the fitted exponent $\gamma_{\alpha}$ as a function of the lower cutoff $L_{\min}$ to reduce finite-size effects while avoiding the instability associated with overly narrow fitting windows. The converged regime is identified using the same criteria as above, and thermodynamic-limit value is then obtained from a linear extrapolation in $1/L_{\min}$. The corresponding results are shown in Fig.~\ref{fig:figs4}.

To determine the scaling form of $F_A$ at the critical point $\alpha_c$, we analyze its system-size dependence over the range $L \in [128,256]$, where finite-size effects are already negligible, as  $L_{min} = 128$ are obtained from the convergence analysis above. Visually, in Fig.~5b of the main text, the data for $F_A$ are approximately linear in a semi-logarithmic plot, which is consistent with a logarithmic dependence on system size. To quantify this observation, we perform weighted fits using the standard errors of $F_A$ as fitting uncertainties and compare a logarithmic ansatz, $F_A = \ell_2^F \ln L + O(1)$, with a power-law form, $F_A = A_{\alpha}L^{\gamma_{\alpha}}$. The logarithmic fit yields a smaller reduced chi-square, $\chi^2_{\mathrm{log}} = 0.53$, than the power-law fit, which gives $\chi^2_{\mathrm{power}} =  6.17$. These results support that $F_A$ follows a logarithmic size dependence at $\alpha_c$, rather than a power-law one. Therefore, in Fig.~5c of the main text, the exponent $\gamma_{\alpha_c}$ is set to zero to indicate the vanishing of the power-law exponent at the transition.

\section{Finite-size scaling and crossing analysis}
\label{sec:supp-fss}

This section gives the analysis underlying the critical point and exponent estimates quoted in Fig.~2 of the main text. The dimensionless Binder cumulant \(U\) and SOP ratio \(R\) are analyzed through their finite-size crossing points for pairs \((L,2L)\). For either observable, the crossing point \(\alpha_*(L)\) is fitted to
\begin{equation}
\alpha_*(L)=\alpha_c+bL^{-(1/\nu+\omega)},
\end{equation}
where \(\omega\) denotes the leading correction-to-scaling exponent. Applying this form separately to \(U\) and \(R\) both give \(\alpha_c=2.49(1)\). The agreement between the two estimates supports the use of a single critical point $\alpha_c=2.49$ in the finite-size collapse analysis below.

For the N\'eel order parameter, we use the finite-size scaling form
\begin{equation}
    m^2L^{2\beta/\nu} = f_m\!\left[\left(\frac{\alpha}{\alpha_c}-1\right)L^{1/\nu} \right],
\end{equation}
with \(\alpha_c=2.49\) fixed from the crossing-point analysis. Near criticality, the scaling function is approximated locally by a cubic polynomial in the scaling variable
\begin{equation}
x=\left(\frac{\alpha}{\alpha_c}-1\right)L^{1/\nu},
\end{equation}
so that
\begin{equation}
m^2(\alpha,L)L^{2\beta/\nu}=c_0+c_1x+c_2x^2+c_3x^3 .
\end{equation}
The collapse is performed using the largest available sizes, \(L=256,320,384,512\) and \(1024\), in the critical window around \(\alpha_c\). For each trial pair \((\beta,\nu)\), the rescaled data are fitted to the cubic form above, and the quality of collapse is quantified by
\begin{equation}
    R^2 = 1-\frac{\sum_i (y_i-\hat{y}_i)^2}{\sum_i (y_i-\bar{y})^2},
\end{equation}
where \(y_i\) denotes the rescaled data, \(\hat{y}_i\) the fitted values, and \(\bar{y}\) the mean of the rescaled data. The optimal exponent pair is identified by maximizing \(R^2\). Statistical uncertainties are estimated by Gaussian resampling of the input data followed by repeating the full collapse procedure. This analysis gives
\begin{equation}
\beta=0.27(2),\qquad \nu=1.71(9)
\end{equation}

\noindent for the N\'eel order parameter. For the uncertainties in the extracted components, we supplement the grid search with a Gaussian-resampling analysis based on the statistical uncertainty of the data. For each data point, we take its sample standard deviation as the width of a normal distribution and generate resampled datasets by adding independent Gaussian noise to the original values. The full collapse analysis is then repeated for each resampled dataset, yielding 1000 estimates of $\beta$ and $\nu$. The quoted uncertainties are taken as the standard deviations of the resulting exponent distributions.

\begin{figure}[htp!]
    \centering
\IfFileExists{figs5.pdf}{\includegraphics[width=0.7\linewidth]{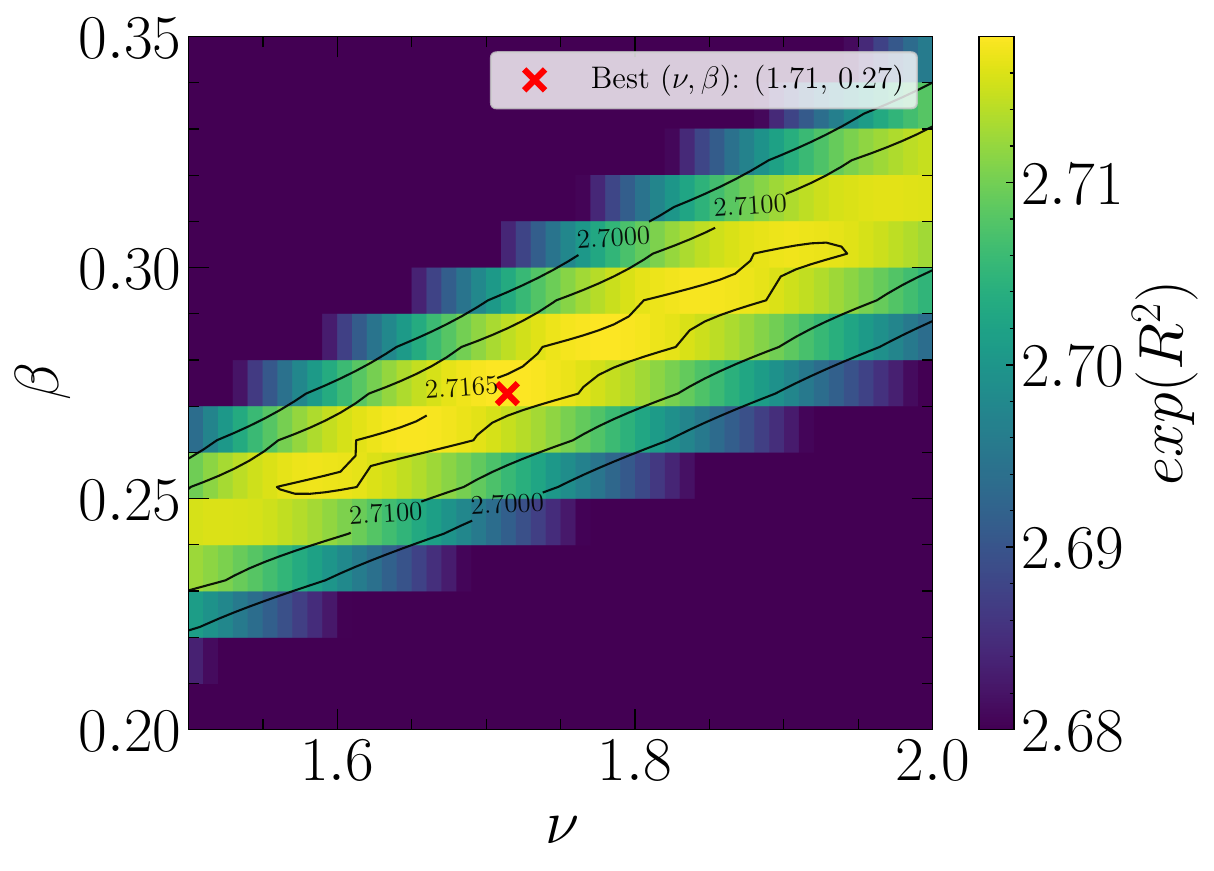}}{%
\fbox{\parbox[c][42mm][c]{0.88\linewidth}{\centering Figure file assumed available for submission: \texttt{figs5.pdf}}}}
    \caption{
    \textbf{Collapse-quality landscape for the critical exponents.} Landscape used to extract the critical exponents \(\beta\) and \(\nu\) from the finite-size scaling of the N\'eel order parameter \(m^2\). The heat map shows the collapse-quality measure \(R^2\) in the \((\beta,\nu)\) plane. The optimal exponent pair, \((\beta,\nu)=(0.27,1.71)\), is determined by maximizing \(R^2\) and corresponds to the main-text collapse shown in Fig.~2c. For visual clarity, the color scale is plotted as \(\exp(R^2)\) to enhance the contrast of the collapse-quality landscape.
    }
    \label{fig:figs5}
\end{figure}

We also apply the same procedure to the string-order structure factor at $\Gamma$, using
\begin{equation}
    S(\Gamma)L^{2\beta/\nu} = f_{\rm SOP}\!\left[\left(\frac{\alpha}{\alpha_c}-1\right)L^{1/\nu}\right] .
\end{equation}
Using the same fixed critical point \(\alpha_c=2.49\), the collapse gives
\begin{equation}
\beta=0.30(8),\qquad \nu=1.74(5).
\end{equation}
These estimates are consistent with those obtained from \(m^2\). In the main text, we therefore quote \(\nu=1.7(1)\) as the representative correlation-length exponent for the subsequent dynamical scaling analysis.

\section{Fourier transform of the string order parameter}
\label{sec:supp-sop-fourier}

In this section, we justify the convention used to define the Fourier transform of the string order parameter (SOP) on a finite periodic chain with system size $L$. Recall the definition of SOP,
\begin{equation}
O^{\text{SOP}}(r) = -\left\langle S_0^z \exp\left(i\pi\sum_{k=1}^{r-1}S_k^z\right) S_r^z \right\rangle
\end{equation}
and in the present work, it is evaluated directly for $r\ge 2 $, where the nonlocal string operator is explicitly defined between the two endpoint spins. However, to perform a discrete Fourier transform on a ring of length $L$, we need $O^{\text{SOP}}(r)$ on a complete set of $L$ consecutive separations, where
\begin{equation}
S(k) = \sum_{r=0}^{L-1} e^{i k r} O^{\text{SOP}}(r), \qquad
k=\frac{2\pi n}{L}, \quad
n=0,1,\dots,L-1,
\end{equation}
which requires a consistent extension to $r=0$ and $r=1$.

\begin{figure}[htp!]
    \centering
\IfFileExists{figs6.pdf}{\includegraphics[width=\linewidth]{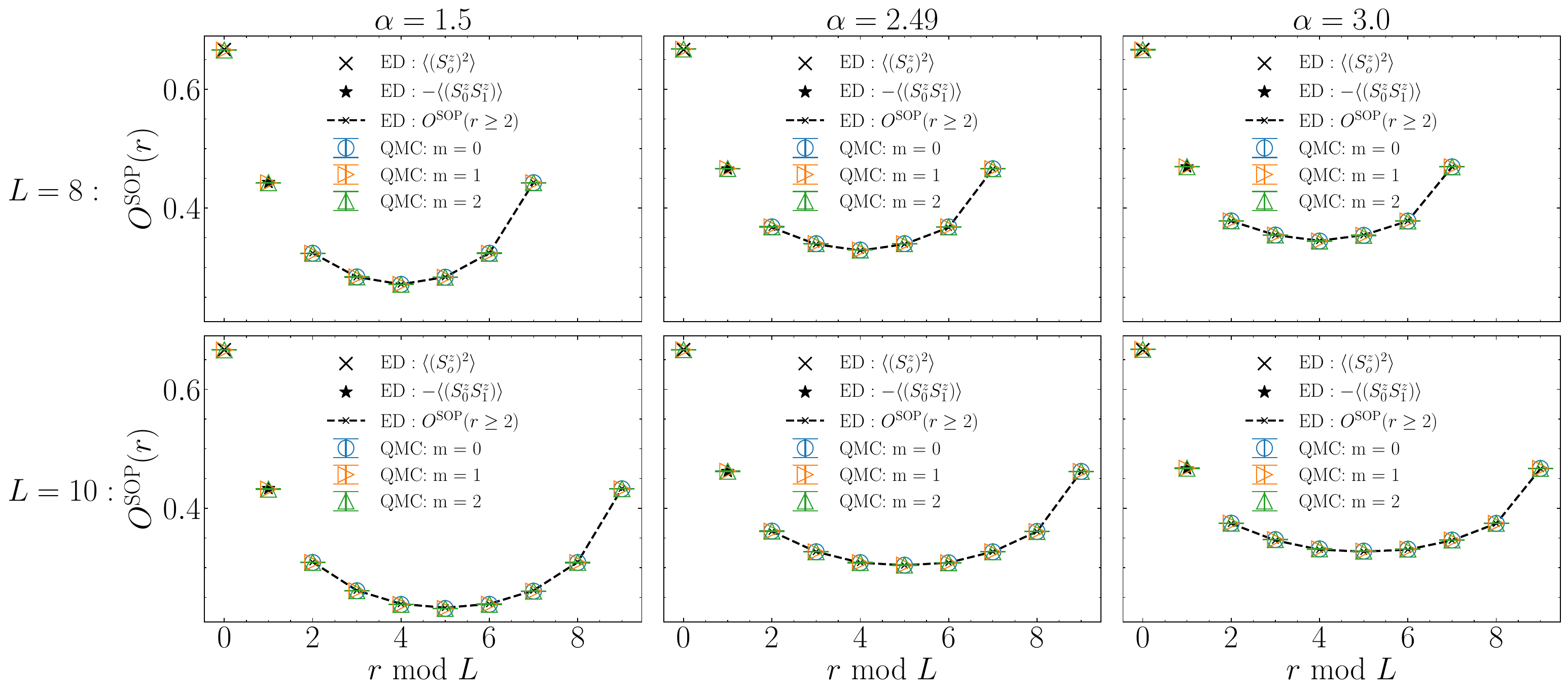}}{%
\fbox{\parbox[c][42mm][c]{0.88\linewidth}{\centering Figure file assumed available for submission: \texttt{figs6.pdf}}}}
    \caption{\textbf{Benchmark of the periodic string-order convention.} Values of $\langle (S_0^z)^2\rangle$, $-\langle S_0^z S_1^z\rangle$ and $O^{\text{SOP}}(r\ge2)$ from exact diagonalization and QMC. Upper panels: Results for $L=8$ in the N\'eel phase ($\alpha = 1.5$), at the critical point ($\alpha = 2.49$), and in the Haldane phase ($\alpha = 3.0$) (left to right). Lower panels: Results for $L=10$ in the N\'eel phase ($\alpha = 1.5$), at the critical point ($\alpha = 2.49$), and in the Haldane phase ($\alpha = 3.0$) (left to right). The results show that $O^{\text{SOP}}(r+mL)=O^{\text{SOP}}(r)$ for $m=1,2$ and $O^{\text{SOP}}(r=mL)$ and $O^{\text{SOP}}(r=mL + 1)$ coincide with $\langle (S_0^z)^2\rangle$ and $-\langle S_0^z S_1^z\rangle$, with $m=1,2$ respectively. }
    \label{fig:figs6}
\end{figure}

This extension is naturally motivated by periodic boundary conditions, $S_{r+L}^z=S_r^z$. In the $\sum_i S_{i}^z=0$ ground-state sector, $O^{\text{SOP}} (r = L) = \langle (S_0^z)^2\rangle$ and $O^{\text{SOP}} (r = L+ 1) = -\langle S_0^z S_1^z\rangle$. These identities motivate the definitions $ O^{\mathrm{SOP}}(0)\equiv \langle (S_0^z)^2\rangle $ and $O^{\mathrm{SOP}}(1)\equiv -\langle S_0^z S_1^z\rangle$. To further verify our extension, we perform QMC simulations and provide exact diagonalization (ED) benchmarking for $L=8$ and $10$. In Fig.~\ref{fig:figs6}, the upper panels show the result for $L=8$ in the N\'eel phase ($\alpha = 1.5$), at the critical point ($\alpha = 2.49$), and in the Haldane phase ($\alpha = 3.0$), and the lower panels show the corresponding results for $L=10$. In each case, the figure compares the ED results for $O^{\text{SOP}}(r\ge2)$, the reference values $\langle (S_0^z)^2\rangle$ and $-\langle S_0^z S_1^z\rangle$, and the QMC simulations for $O^{\text{SOP}}(r)$ over $r\in[2,3L-1]$. As shown in Fig.~\ref{fig:figs6}, the QMC data are periodic with period $L$ and agree with the ED results wherever they overlap, consistent with $O^{\text{SOP}}(r+mL)=O^{\text{SOP}}(r)$ for $m=1,2$. Moreover, the $O^{\text{SOP}}(r=mL)$ and  $O^{\text{SOP}}(r=mL + 1)$ coincide with $\langle (S_0^z)^2\rangle$ and $-\langle S_0^z S_1^z\rangle$, respectively, with $m = 1,2$. We therefore adopt
\begin{equation}
O^{\text{SOP}}(r = 0)= \langle (S_0^z)^2\rangle, \qquad
O^{\text{SOP}}(r = 1)=-\langle S_0^z S_1^z\rangle
\end{equation}
when evaluating the Fourier transform of the SOP.

\section{Spin-wave analysis}
\label{sec:supp-spin-wave}
Linear spin-wave (SW) theory has been developed for the long-range Heisenberg and XXZ chain in Refs.~\cite{yusufSpin2004,laflorencieCritical2005,frerotEntanglement2017}. Because of the long-range nature of the couplings, the SW dispersion relation is sublinear at low momentum $k$ of the form
\begin{equation}
\omega(k)\propto |k|^z ,
\label{eq:sw-low-k}
\end{equation}
with spin-wave dynamical exponent \(z_{\rm SW}=(\alpha-1)/2\) for \(\alpha<3\). This ordered-phase spin-wave exponent should be distinguished from the critical gap exponent \(z=0.81(1)\) extracted at \(\alpha_c=2.49\) in the main text. Consequently, the \(1/S\) correction to the order parameter,
\begin{equation}
\Delta M_{\rm AF}\propto \int \frac{\mathrm{d}k}{\omega(k)} ,
\label{eq:sw-order-correction}
\end{equation}
can remain finite for decay exponents \(\alpha<3\).

As shown in Ref.~\cite{laflorencieCritical2005}, the SW dispersion can be written as
\begin{equation}
 \omega(k) = \sqrt{\left[\gamma - f(k)
  \right]^2 - [g(k)]^2 } ,
\end{equation}
with
\begin{equation}
\begin{aligned}
\gamma &= 2\sum_{n=1}^{\infty}\frac{1}{(2n-1)^\alpha},\\
f(k) &= 2\sum_{n=1}^{\infty}\frac{\cos(2kn)-1}{(2n)^\alpha},\\
g(k) &= 2\sum_{n=1}^{\infty}\frac{\cos[k(2n-1)]}{(2n-1)^\alpha}.
\end{aligned}
\label{eq:sw-functions}
\end{equation}

\begin{figure}[htbp]
\centering
\IfFileExists{figs7.pdf}{\includegraphics[width=0.6\linewidth]{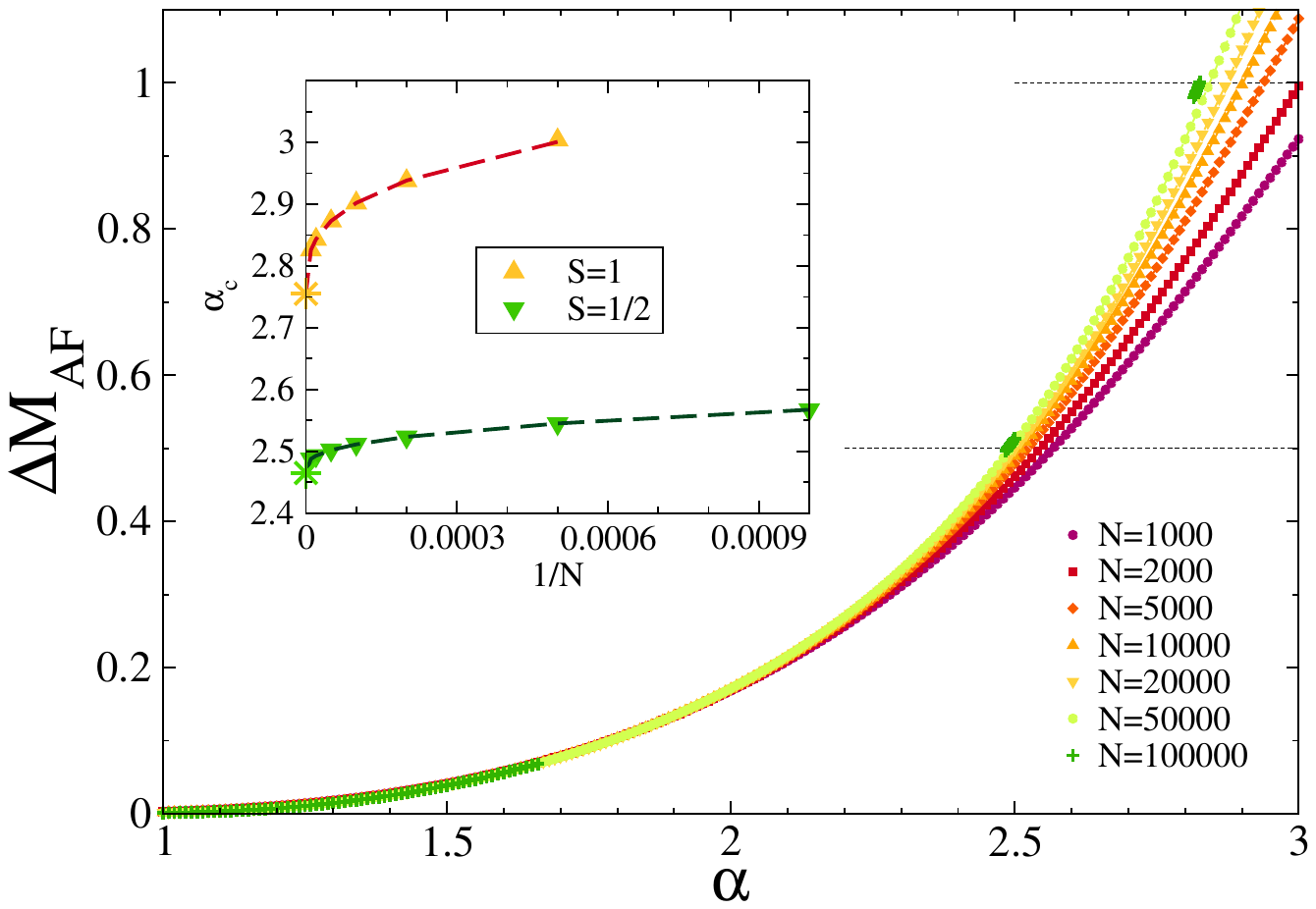}}{%
\fbox{\parbox[c][42mm][c]{0.6\linewidth}{\centering Figure file assumed available for submission: \texttt{figs7.pdf}}}}
\caption{\textbf{Linear spin-wave results for the long-range Heisenberg model.} The main panel shows the spin-wave correction to the N\'eel order parameter Equation~(\ref{eq:DeltaMAF}), evaluated using a finite sum over discretized momentum $k_{\ell}=\frac{\pi}{N}(\ell-\frac{1}{2})$ with $\ell=1,\ldots,N/2$ for various values of $N$, as indicated on the plot. The inset shows how the intercept $\Delta M_{\rm AF}=S$ converges to $\alpha_c^{\rm SW}\approx 2.46$  for $S=1/2$ and to $\alpha_c^{\rm SW}\approx 2.75$ for $S=1$  when $1/N\to 0$.\label{fig:figSW}}
\end{figure}

 The $1/S$ correction to the staggered magnetization is given by the following integral~\cite{laflorencieCritical2005}
\begin{equation}
\Delta M_{\rm AF} = \frac{1}{2\pi}\int_{-\pi/2}^{\pi/2}\mathrm{d}k
\left[\frac{\gamma-f(k)}{\omega_k}-1\right],
\label{eq:DeltaMAF}
    \end{equation}
that we approximate with a finite sum over discretized momentum $k_{\ell}=\frac{\pi}{N}(\ell-\frac{1}{2})$ with $\ell=1,\ldots,N/2$. The obtained sum converges slowly with the number of points $N$, as shown in Fig.~\ref{fig:figSW}.

The condition  $\Delta M_{\rm AF}= S$  allows us to find the SW estimates for the value
of the critical decay $\alpha_{c}^{sw}$ below which long-range N\'eel order can exist. For $S=1/2$ we recover the value $\approx 2.46$ reported in Ref.~\cite{laflorencieCritical2005} and for $S=1$  we find $\alpha_{c}^{sw} (S=1)\approx 2.75$. In both cases, the relative error with respect to the exact QMC estimate is $\frac{\alpha_c^{\rm SW}-\alpha_c^{\rm QMC}}{\frac{1}{2}(\alpha_c^{\rm SW}+\alpha_c^{\rm QMC})}\approx 10\%$

\section{Critical exponents and hyperscaling relations}
\label{sec:supp-hyperscaling}
Here we briefly recall the zero-temperature quantum critical properties of the order parameter and various physical quantities that are governed by the set of (quantum) critical exponents $\beta$, $\nu$, $z$, and $\eta$.
\subsection{N\'eel order parameter and critical exponents}
The (square) N\'eel order parameter
\begin{equation}
M^2_{\rm AF}=\lim_{L\to \infty} {3\left\langle\left(\frac{1}{L} \sum_{i=1}^{L} (-1)^i S_i^z\right)^2\right\rangle}
\end{equation}
is expected to obey the following critical behavior when $\alpha_c$ is approached from below
\begin{equation}
    M^2_{\rm AF}\propto (\alpha_c-\alpha)^{2\beta}.
\end{equation}
An equivalent definition of the order parameter relies on the long distance spin-spin correlation function~\cite{sandvik_finite_1997}
\begin{equation}
    C^z(r)=(-1)^r\langle S_i^zS_{i+r}^z\rangle,
\end{equation}
such that
\begin{equation}
M^2_{\rm AF}=\lim_{r\to \infty} {3C^z(r)}.
\end{equation}
At criticality, the critical  decay is algebraic
\begin{equation}
    C^z(r)\propto \frac{1}{{r}^{d+z+\eta-2}},
    \label{eq:Cz}
\end{equation}
where $d$ is the dimension (here $d=1$), $z$ the dynamical exponent, and $\eta$ the anomalous exponent.

The finite-size scaling of the gap at criticality is controlled by the dynamical exponent
\begin{equation}
    \Delta(L)\propto L^{-z},
\end{equation}
which upon approaching the QCP from the disordered gapped regime yields
\begin{equation}
    \Delta\propto (\alpha-\alpha_c)^{\nu z}.
\end{equation}

The anomalous exponent can also be obtained from the size-dependence of the zero-temperature (staggered) susceptibility (associated to the antiferromagnetic instability)~\cite{laflorencieCritical2005}
\begin{equation}
    \chi_{\rm stag}(L)\propto L^{2-\eta}.
\end{equation}
For $S=1/2$, Ref.~\cite{laflorencieCritical2005} reported very good agreement between the QMC estimate of $\eta$ and the  simple mean-field prediction~\cite{sak_recursion_1973}
\begin{equation}
    \eta=3-\alpha
\end{equation}

\subsection{Hyperscaling relations}

In the close vicinity of the critical point $\alpha\to\alpha_c$, one can replace $r$ by the correlation length in Equation~(\ref{eq:Cz})
\begin{equation}
    \xi\sim (\alpha_c-\alpha)^{-\nu},
\end{equation}
which yields the usual hyperscaling relation in $d=1$
\begin{equation}
2\beta=\nu(z+\eta-1).
\label{eq:hsc}
\end{equation}
We report in Table~\ref{tab:1} below the exponents for the QLRO-to-N\'eel transition for $S=1/2$~\cite{laflorencieCritical2005,zhaoUnconventional2025}, and the Haldane-to-N\'eel for the present $S=1$ study. The validity of the hyperscaling relation Equation~(\ref{eq:hsc}) is also checked in the last column.
\begin{table}[htbp]
  \centering
\resizebox{\textwidth}{!}{%
\begin{tabular}{l|c|c|c|c|c|c}
 & $\alpha_c$ & $\beta$ & $\nu$ & $z$ & $\eta$ &$2\beta-\nu(z+\eta-1)$\\
\hline
$S = {1}/{2}$ (Refs.~\cite{laflorencieCritical2005,zhaoUnconventional2025})& 2.231(1) & 0.57(2) &2.16(1)& 0.75(5) & 0.77(2)&0.0168\\

$S = 1$ (This work) & 2.49(1) & 0.29(10)& 1.7(1)& 0.81(1) & $3-\alpha_c$ & $0.036$\\
\bottomrule
\end{tabular}%
}
\caption{\textbf{Critical parameters of the long-range spin models} Parameters for \(S=1/2\) from Refs.~\cite{laflorencieCritical2005,zhaoUnconventional2025} and for \(S=1\) from this work. The final column tests Equation~(\ref{eq:hsc}). For \(S=1\), the anomalous exponent was not fitted independently; the estimate \(\eta=3-\alpha_c\) was used~\cite{sak_recursion_1973}.\label{tab:1}}
\end{table}

\end{appendix}

\bibliography{main_scipost}
\end{document}